\documentstyle[epsf,epsfig,12pt]{article}
\mark{{}{}}
\def\nll{ \nonumber \\}
\def\ti{t_{_1}}
\def\tii{t_{_2}}
\def\tiii{t_{_3}}
\def\tiv{t_{_4}}
\def\lb{\left(}
\def\rb{\right)}
\def\tw{t_{_W}}
\def\gl{g_{_L}}
\def\gr{g_{_R}}
\def\psla{\hbox{\rlap/p}}
\def\qsla{\hbox{\rlap/q}}
\def\nsla{\hbox{\rlap/n}}
\def\lsla{\hbox{\rlap/l}}
\def\msla{\hbox{\rlap/m}}
\def\cnsla{\hbox{\rlap/N}}
\def\clsla{\hbox{\rlap/L}}
\def\cmsla{\hbox{\rlap/M}}
\def\z0{Z}
\def\gf{G_{\mu}}
\def\zm{M_{_Z}}

\def\msb{{\overline{MS}}}

\def\hm{M_{_H}}
\def\wm{M_{_W}}

\def\gw{\Gamma_{_W}}
\def\sgw{\gamma_{_W}}
\def\muw{\mu_{_W}}

\def\barf{\overline f}

\def\barb{\overline b}
\def\bard{\overline d}

\def\baru{\overline u}
\def\barv{\overline v}
\def\barnu{\overline{\nu}}

\def\shat{\hat s}

\def\i3f{I^{(3)}_f}

\def\osp2{16\,\pi^2}
\def\ap2{\left(p^2\right)}

\def\s0h{\sigma^h_0}
\def\ba{\begin{eqnarray}}
\def\ea{\end{eqnarray}}

\def\beq{\begin{equation}}
\def\eeq{\end{equation}}
\def\bea{\begin{eqnarray}}
\def\eea{\end{eqnarray}}
\def\barr{\begin{array}}
\def\earr{\end{array}}
\def\bc{\begin{center}}
\def\ec{\end{center}}
\def\btab{\begin{tabular}}
\def\etab{\end{tabular}}

\def\veps{\varepsilon}

\begin{document}

\title{\bf WTO - A DETERMINISTIC APPROACH\\
TO $4$-FERMION PHYSICS}
\author{
Giampiero PASSARINO$^{ab}$,
}
\date{}

\maketitle

\begin{itemize}

\item[$^a$]
             Dipartimento di Fisica Teorica,
             Universit\`a di Torino, Torino, Italy
\item[$^b$]
             INFN, Sezione di Torino, Torino, Italy

\end{itemize}

\noindent
email:
\\
giampiero@to.infn.it

\vspace{4cm}
The program WTO, which is designed for computing cross sections and other
relevant observables in the $e^+e^-$ annihilation into four fermions, is
described. The various quantities are computed over both a completely 
inclusive experimental set-up and a realistic one, i.e. with cuts on the final
state energies, final state angles, scattering angles and final state
invariant masses. Initial state QED corrections are included by means of the 
structure function approach while final state QCD corrections are applicable
in their {\it naive} formulation. A gauge restoring mechanism is included
according to the Fermion-Loop scheme. The program structure is highly modular 
and particular care has been devoted to computing efficiency and speed.

\newpage
\tableofcontents

\section{Program summary}
\noindent
{\it Title of program:} WTO
\vskip 15pt

\noindent
{\it Computer:} Alpha AXP-2100 (VAXstation$\,\cdot\,$4000$\,\cdot\,$90); {\it 
Installation:} INFN, Sezione di Torino, via P.Giuria 1, 10125 Torino, Italy
\vskip 15pt

\noindent
{\it Operating system:} OpenVMS(VMS)
\vskip 15pt

\noindent
{\it Programming language used:} FORTRAN 77
\vskip 15pt

\noindent
{\it Memory required to execute with typical data:} Peak working set size:
3000-3500
\vskip 15pt

\noindent
{\it No. of bits in a word:} 64(32)
\vskip 15pt

\noindent
{\it No. of lines in distributed program:} 35313
\vskip 15pt

\noindent
{\it subprograms used:} NAGLIB~\cite{nag}
\vskip 15pt

\noindent
{\it Keyords:} $e^+e^-$ annihilation into four fermions, LEP~2, properties
of the $W$ vector boson, Higgs boson, initial and final state QED radiation, 
QCD corrections, Minimal Standard Model, cross sections and moments of 
distributions, gauge invariance, theoretical error, deterministic integration.
\vskip 15pt

\noindent
{\it Nature of physical problem} An accurate description of the process
$e^+e^- \to 4\,$fermions is needed in order to fully describe the physics
available at LEP~2 and at higher energies. In particular the properties
of the $W$ boson can be correctly analyzed around and above the threshold
only when the full gauge-invariant set of diagrams contributing to a given
final state is included. Similarly the $Z-Z$ production and the Higgs
boson production can be studied within the minimal standard model.
Indeed from a field theoretical point of view both the $W$ and the $Z$
bosons (and the Higgs boson too) are unstable particles and a full
description of their production can only proceed through the complete set
of matrix elements for the $4$-fermion processes. Although at typical LEP~2
energies the difference between the full calculation and the {\it double
resonant} approximation for a process like $e^+e^- \to \mu^-\barnu_{\mu}
\nu_{\tau}\tau^+$ is totally negligible, the same is certainly not true
anymore when we consider $e^+e^- \to e^-\barnu_e\nu_{\mu}\mu^+$ or
any of the neutral current processes.

\vskip 15pt

\noindent
{\it Method of solution} The helicity amplitudes for each given process
are given, according to the formalism of ref.~\cite{mhf}, in terms of
the $7$ independent invariant which characterize the phase space.
The phase space itself, including all realistic kinematical cuts, is also
described in terms of invariants. Initial state QED radiation is included
by means of the structure function approach. Upon initialization the final
state QCD corrections are included by adopting a {\it naive} approach
(NQCD). The numerical integration, with complete cut-availability, is
performed with the help of a {\it deterministic} integration routine
which makes use of quasi-random, deterministic number sets, the shifted
Korobov sets. The boundaries of the phase space, with kinematical cuts,
are reconstructed through a backwards propagation of constraints.
\vskip 15pt

\noindent
{\it Restrictions on the complexity of the problem} The theoretical
formulation is specifically worked out for massless fermions, although this
is not a limitation of principle. Emission of photons is strictly collinear 
and no $p_t$ is therefore included. No interface exists with the standard 
packages for {\it hadronization}.
\vskip 15pt

\noindent
{\it Typical running time} Dependent on the process and on the required
accuracy. For example a $0.1\%$ accuracy for a CC11 process requires 
something of the order of $440$ CPU seconds for Alpha AXP-2100. On the other
extreme a $0.5\%$ accuracy for some (but not all) of the NC processes 
requires approximately $12$ CPU hours on the same computer.

\section{Long Write-up}

\subsection{Introduction}

WTO is a {\it quasi-analytical, deterministic} code for computing
observables related to the process

\begin{equation}
e^+e^- \to f_1\barf_2 f_3\barf_4.
\end{equation}

\noindent
If one neglects the fermion masses there are $32$ distinct processes of this 
kind (classified in ref.~\cite{wweg}),
$29$ of which are at present accessible with WTO. The only exclusion is
at present given by $M^{\dagger}M$ with $M = E\otimes E, E=(\nu_e,e)$. 

We have 
several codes, documented in the literature~\cite{lep2}-\cite{yr}-\cite{vwg}, 
which perform a 
similar task and all of them can be classified into three broad families, i.e. 
semi-analytical, MonteCarlo (MC) integrators or classical event generator. A 
semi-analytical code will perform as much as possible of the $7$($9$ with 
initial state QED radiation included through structure function approach) 
phase space integrations leaving at most $2(3)$ of them for some kind of 
numerical evaluation. The advantage is represented by computational speed and
high numerical accuracy but only very few and selected cuts can be applied,
typically on two of the final state invariant masses. From LEP~1 experience
we know that semi-analytical calculations are relevant also to
experimentalists and not only as the ideal source of benchmark.
A MC approach clearly represents the other extreme giving full cut-availability
but with no intermediate analytical step taking place in the process of
obtaining the final result out of the matrix elements. 
Kinematical cuts are, to a large extent, treated with the help of {\tt IF 
statements} and the efficiency of the integration procedure is dominated by 
the {\it adaptivity} of the integration routine.

WTO has been designed to lay somehow in between these two approaches.
Basically the matrix elements are generated through the helicity
formalism of ref.~\cite{mhf} and they are compact expressions completely
given in terms of the invariants which describe the process. Also the
momenta of the final states are, component by component, given in terms
of the invariants used in the integration over the phase space, thus allowing
to implement the kinematical cuts with an analytical control
(almost completely). There are very few parts of the code where this is not
implemented, having to do with the fact that in any $2 \to 4$ process
there is a non-linear constraint among the invariants. Although this
is not a limitation of principle we have nevertheless verified that
with an analytical treatment of the non-linear constraint
the balance between CPU time and efficiency in the integration is not in 
favor of the former and, for this reason, some numerical control of the phase 
space boundaries has survived in WTO. 

Having explained the {\it quasi-analytical} treatment of the calculation in 
WTO we spend now few words on the {\it deterministic} approach to the
integration. The integration over the phase space is performed in WTO with 
the help of the NAG~\cite{nag} routine D01GCF. This routine uses the 
Korobov-Conroy number theoretic approach with a MC error estimate arising 
from converting the number theoretic formula for the $n$-cube $[0,1]^n$ into 
a stochastic integration rule. This allows a `standard error' to be estimated. 
There is no adaptive strategy at work since the routine D01GCF,
being a {\it deterministic} one, will use a fixed grid. The evaluation of the
specified observable will be repeated {\tt NRAND} times to give the final 
answer, however there is no possibility to examine the partial results but 
only the average and the resulting standard error will be printed. The error in 
evaluating, say a cross section, satisfies $E < CK\,p^{-\alpha}
\log^{\alpha\beta}p$, where $p=${\tt NPTS}, $\alpha$ and $C$ are real numbers 
depending on the convergence rate of the Fourier series, $\beta$ is a constant 
depending on the dimensionality $n$ of the integral and $K$ is a constant 
depending on $\alpha$ and $n$.
        
\subsection{The phase space}

The process under consideration is specified by

\begin{equation}
e^+(p_+)e^-(p_-) \to f_1(q_1) \barf_2(q_2) f_3(q_3) \barf_4(q_4),
\end{equation}

\noindent
and it is described in terms of $14$ invariants (plus the c.m.s. energy)

\begin{eqnarray}
s &=& -(p_+ + p_-)^2, \nll
x_{1i}s &=& -(p_+ + q_{i-2})^2,\qquad i=3,\dots,6, \nll
x_{2i}s &=& -(p_- + q_{i-2})^2,\qquad i=3,\dots,6, \nll
x_{ij}s &=& -(q_{i-2} + q_{j-2})^2,\qquad i\not=j=3,\dots,6.
\end{eqnarray}

\noindent
Out of these we select $7$ linearly independent combinations of invariants,
i.e.

\begin{equation}
m_-^2,\quad m_+^2,\quad M_0^2,\quad m_0^2,\quad m^2,\quad \tw,\quad \ti,
\label{inv}
\end{equation}

\noindent
such that

\begin{eqnarray}
x_{_{13}} &=& \ti,\quad x_{_{14}} = \tw-\ti,\quad x_{_{15}} = \tiii,
\quad x_{_{16}} = 1-\tw-\tiii, \nll
x_{_{23}} &=& m_-^2+m_0^2+m^2-\ti,\quad x_{_{24}} = 1-m_+^2-m_0^2-m^2-\tw+\ti, 
\nll
x_{_{25}} &=& m_+^2+M_0^2+m^2-\tiii,\quad x_{_{26}} = -m_-^2-M_0^2-m^2+\tw+\tiii, 
\nll
x_{_{34}} &=& m_-^2,\quad x_{_{35}} = m^2,\quad x_{36} = m_0^2,\nll
x_{_{45}} &=& M_0^2,\quad x_{_{46}} = 1-m_-^2-m_+^2-M_0^2-m_0^2-m^2, \nll
x_{_{56}} &=& m_+^2.
\end{eqnarray}

\noindent
An auxiliary variable has been introduced, $\tiii$, which is fixed by the 
non-linear constraint. Additional auxiliary variables to be considered
are also

\begin{eqnarray}
e_1 &=& m_-^2+m_0^2+m^2,\quad e_2 = 1+m_-^2-m_+^2-e_1, \nll
e_3 &=& m_+^2+M_0^2+M^2,\quad e_4 = 1+m_+^2-m_-^2-e_3, \nll
\tii &=& \tw-\ti,\quad \tiv = 1-\tw-\tiii.
\end{eqnarray}

\noindent
This choice is far from being arbitrary and reflects the need of
selecting those combinations of invariants which characterize the peak
structure of the integrand.
With the help of the previous quantities we can reconstruct completely the phase
space. If $M_{ij}, E_i, \theta_i$ and $\psi_{ij}$ (i,j=1,4) are the final
state invariant masses, the final state energies, the scattering angles and
the final state angles respectively, then in the $e^+e^-$ c.m.s. we can write

\begin{eqnarray}
M^2_{12} &=& m_-^2,\quad M^2_{13} = m^2,\quad M^2_{14} = m_0^2, \nll
M^2_{23} &=& M_0^2,\quad M^2_{24} = 1-m_-^2-m_+^2-M_0^2-m_0^2-m^2, \nll
M^2_{34} &=& m_+^2, 
\label{nrps}
\end{eqnarray}

\noindent
and

\begin{eqnarray}
E_i &=& \frac{1}{2}\, e_i\,\sqrt{s}, \qquad i=1,\dots,4, \nll
\cos\theta_i &=& 1 - {{t_i}\over {e_i}}, \qquad i=1,\dots,4, \nll
\cos\psi_{12} &=& 1 - 2\,{{m_-^2}\over {e_1e_2}}, \nll
\cos\psi_{13} &=& 1 - 2\,{{m^2}\over {e_1e_3}}, \qquad {\hbox{etc.}}
\end{eqnarray}

\noindent
Clearly we want to take into account the QED radiation from initial states
which is done through the formalism of the structure 
function~\cite{sf1}-\cite{sf3}. 
In this way the {\it hard} process becomes

\begin{equation}
e^+(x_+p_+)e^-(x_-p_-) \to f_1(q_1) \barf_2(q_2) f_3(q_3) \barf_4(q_4),
\end{equation}

\noindent
and the corresponding kernel cross section is afterwards convoluted with 
the standard structure functions. Since we want to impose kinematical cuts
in the laboratory frame, all the quantities must be reconstructed in presence
of e.m. radiation. We have

\begin{eqnarray}
E_i &=& \frac{1}{2}\,\lb x_+e_i - \Delta t_i\rb \,\sqrt{s}, \qquad
\Delta = x_+ - x_-, \nll
\cos\theta_i &=& 1 - 2\,{{x_-t_i}\over {x_+e_i - \Delta t_i}}, \nll
\cos\psi_{12} &=& 1 - \frac{1}{2}\,{{m_-^2s}\over {E_1E_2}}, \quad {\hbox{etc.}}
\label{ps}
\end{eqnarray}

\noindent
Formally any distribution is written as

\begin{equation}
F = \int\,dx_+dx_-\,\int\,dPS\,\Theta_{cut} D(x_+,s),D(x_-,s)\,f,
\end{equation}

\noindent
where $f$ refers to the kernel-distribution and $D$ to the structure function. 
Thus $f$ will depend on

\begin{eqnarray}
x_+p_+\cdot q_i &=& - \frac{1}{2}\,x_+x_{1,i+2}s, \nll
x_-p_-\cdot q_i &=& - \frac{1}{2}\,x_-x_{2,i+2}s, \nll
q_i\cdot q_j &=& - \frac{1}{2}\,x_+x_-x_{i+2,j+2}s.
\end{eqnarray}

\noindent
In presence of radiation we write

\begin{eqnarray}
x_{_{34}} &=& x_+x_-m_-^2, \dots , x_{_{56}} = x_+x_-m_+^2, \nll
x_{_{13}} &=& x_-\ti, \quad etc., \nll
x_{_{23}} &=& x_+\lb e_1-\ti\rb , \quad etc.
\end{eqnarray}

\noindent
Consequently

\begin{equation}
x_+p_+\cdot q_1 = -\frac{1}{2}x_+x_{_{13}}s = -\frac{1}{2}x_+x_-\ti s =
- \frac{1}{2}\ti\shat, \quad {\hbox{etc.}},
\end{equation}

\noindent
where $\shat = x_+x_-s$.
$f$ can be given in terms of the invariants $\{I\} = m_-^2,\dots,\ti$
if we normalize everything with respect to $\shat$. Indeed we obtain

\begin{equation}
{{d^7F}\over {dm_-^2\dots d\ti}} = \int\,dx_+dx_-\,\Theta_{cut}
D(x_+,s)D(x_-,s)\,f\lb \{I\},\shat\rb \,\int\,dPS\,\delta_I,
\end{equation}

\noindent
where $\delta_I$ is the collection of delta-functions giving the invariants
$\{I\}$ in terms of scalar products.

We now come back to the auxiliary invariant $\tiii$. The easiest way of
obtaining the non-linear constraint is to start in the c.m.s of
$P_{\pm} = x_{\pm}p_{\pm}$ with $P_{\pm}$ along the $z$-axis. Then
$q_1$, in the $x-z$ plane, is constructed, component by component in terms
of invariants. Obviously $q_4$ is fixed by energy-momentum conservation
and $q_2, q_3$ will be constructed recursively but for $q_3$ we allow
for a fifth component. The request that this extra component be zero
gives the non-linear constraint and therefore $\tiii$ in terms of the
remaining independent invariants. The non-linear constraint, being
quadratic in $\tiii$, gives rise to two distinct solutions which must be
properly interpreted. The phase space integral can be written as

\begin{equation}
\int\,dPS = \int\,\Pi_i\,d^4q_i\,\delta^+\lb q_i^2\rb \delta^4
\lb P-\sum_iq_i\rb ,
\end{equation}

\noindent
with $P^2 = - \shat$. After some manipulations we obtain

\begin{eqnarray}
\int\,dPS &=& \frac{1}{8}\shat^{3/2}\int\,\Pi_ide_i\Pi_jdt_jdm_-^2dm^2\,
\Pi_i\theta(e_i)\theta(2-\sum_ie_i), \nll
 & & \mbox{} \times\int\,\Pi_id^3q_i\,\delta\lb q_i^2-\frac{1}{4}e_i^2\shat\rb 
\delta(X_1)\delta(X_2)\delta(X_{12})\,J,
\end{eqnarray}

\noindent
where

\begin{eqnarray}
X_1 &=& \ti + \frac{1}{2}(c_1-1)e_1, \nll
X_2 &=& \tii + \frac{1}{2}(c_2-1)e_2, \nll
X_{12} &=& m_-^2 + \frac{1}{2}(c_{12}-1)e_1e_2.
\end{eqnarray}

\noindent
Here $c_i$ denotes $\cos\theta_i$ in the c.m.s. (with $P_+$ along the 
positive $z$-axis) and $c_{12}$ is the cosine of the angle between 
$q_1$ and $q_2$ in the same system. Moreover

\begin{eqnarray}
J &=& \int\,d^3q_3\,\delta\lb q_3^2 - \frac{1}{4}e_3^2\shat\rb \,
\delta\lb m^2 + \frac{1}{2}(c_{13}-1)e_1e_3\rb  \nll
 & & \mbox{}\times \delta\lb \sum_ie_i - m_-^2 - m^2 + \frac{1}{2}(c_{23}-1)
e_2e_3\shat\rb .
\end{eqnarray}

\noindent
To compute $J$ we choose now another reference frame ($II$) given by

\begin{eqnarray}
q_1 &=& \frac{1}{2}e_1\sqrt{\shat}\lb 0,0,1,1\rb , \nll
q_2 &=& \frac{1}{2}e_2\sqrt{\shat}\lb s_{12},0,c_{12},1\rb , \nll
q_3 &=& \frac{1}{2}e_3\sqrt{\shat}\lb s_{13}\cos\phi,s_{13}\sin\phi,
c_{13},1\rb .
\end{eqnarray}

\noindent
Let $\Phi = s_{13}s_{12}\cos\phi + c_{13}c_{12} -1$ computed for
$c_{13} = 1 - 2m^2/e_1e_3$. We get

\begin{eqnarray}
J &=& {1\over {2\,e_1\sqrt{\shat}}}\,\left[\int_0^{\pi}\,d\phi +
\int_{\pi}^{2\,\pi}\,d\phi\right]  \nll
 & & \mbox{}\times \delta\lb \lb \sum_ie_i - m_-^2 - m^2 - 1 + \frac{1}{2}
\Phi\rb \shat\rb   \nll
 &=& {1\over {2\,\sqrt{\shat}}}\,{{\theta_> + \theta_<}\over R_{_4}}.
\end{eqnarray}

\noindent
Before carrying on the full integration we can now connect the two
$\theta$ functions, $\theta_>$ and $\theta_<$, with the non-linear constraint.
Let $\tiii^{\pm}$ be the two solutions and moreover let
$q_3^{\pm}(I)$ be $q_3$ in the c.m.s computed according to the
choice $\tiii = \tiii^+$ or $\tiii = \tiii^-$. Let finally $q_3(II)$ be $q_3$ 
as given in system $II$ with

\begin{equation}
c_{13} = 1 - 2\,{{m^2}\over {e_1e_3}}, \qquad s_{13}^2 = 1 - c_{13}^2.
\end{equation}

\noindent
To fix $\cos\phi$ we first transform $q_3(I)$ from system $I$ to
system $II$ and compare $q_3^t(I)$ with $q_3(II)$. The root for $\tiii$
is chosen to match the two vectors, one root will correspond to $\sin\phi>0$
($\theta_>$) and the other to $\sin\phi<0$ ($\theta_<$) for fixed $\cos\phi$.
Moreover

\begin{eqnarray}
R^2_{_4} &=& m^2e_2^2\lb e_1e_3 - m^2\rb  s_{12}^2  \nll
 & & \mbox{} - \left[ \frac{1}{2}e_2\lb e_1e_3 - 2\,m^2\rb c_{12} + e_1\lb
\sum_ie_i-m_-^2-m^2-1-\frac{1}{2}e_2e_3\rb\right]^2.
\end{eqnarray}

\noindent
where

\begin{equation}
c_{12} = 1 - 2\,{{m_-^2}\over {e_1e_2}}.
\end{equation}

\noindent
Thus

\begin{eqnarray}
\int\,dPS &=& \frac{1}{8}\shat\,\int\,d\{I\}\,\Pi_i\theta(e_i)
\theta\lb 2-\sum_ie_i\rb\theta\lb R^2_{_4}\rb\lb\theta_>+\theta_<\rb  \nll
 & & \mbox{} \times \int\,d^3q_1\,\delta\lb q_1^2-\frac{1}{4}e_1^2\shat\rb
\delta\lb\ti+\frac{1}{2}e_1\lb(c_1-1\rb\rb\,H,
\end{eqnarray}

\noindent
where $\{I\}$ denotes collectively all the invariants and

\begin{eqnarray}
H &=& R^{-1}_{_4}\,\int\,d^3q_2\,\delta\lb q_2^2-\frac{1}{4}e_2^2\shat\rb\delta
\lb \ti+\tw+\frac{1}{2}e_2\lb c_2-1\rb\rb  \nll
 & & \mbox{} \times \delta\lb m_-^2+\frac{1}{2}e_1e_2\lb c_{12}-1\rb\rb.
\end{eqnarray}

\noindent
As a final step we compute $H$ by going back to system I. Here

\begin{eqnarray}
q_1 &=& \frac{1}{2}e_1\sqrt{\shat}\lb s_1,0,c_1,1\rb,  \nll
q_2 &=& \frac{1}{2}e_1\sqrt{\shat}\lb s_2\cos\psi,s_2\sin\psi,c_2,1\rb,
\end{eqnarray}

\noindent
and

\begin{equation}
H = {{\sqrt{\shat}}\over {R_{_2}R_{_4}}}.
\end{equation}

\subsection{Boundaries of the phase space}

In this section we consider the problem of {\it reconstructing} the full
phase space in terms of our independent invariants.
There are {\it natural} limits on our integration variables which can be
derived by clustering the final state momenta $q_1,\dots,q_4$ into pairs,
i.e. $(q_1+q_2),(q_3+q_4)$ or $(q_1+q_3),(q_2+q_4)$ or $(q_1+q_4),(q_2+q_3)$ 
and by imposing the phase space constraints relative to a $2 \to 2$
process. To give an example we introduce

\begin{equation}
P_{\pm} = x_{\pm}p_{\pm}, \qquad Q_- = q_1+q_2, \quad Q_+ = q_3+q_4.
\end{equation}

\noindent
We find

\begin{eqnarray}
\shat &=& - \lb P_+ + P_-\rb^2, \nll
{\hat t} &=& - \lb P_+ - Q_-\rb  = \lb m_-^2 - \tw\rb^2\shat.
\end{eqnarray}

\noindent
The condition to be fulfilled is that ${\hat t}$ must be in the physical
region, therefore $X \geq  0$ with $X$ given by

\begin{equation}
X = - \tw^2 + \lb 1 + m_-^2 + m_+^2\rb \tw - m_-^2.
\end{equation}

\noindent
This fixes the {\it natural} limits of the variable $\tw$, i.e.

\begin{eqnarray}
T_- &\leq & \tw \leq  T_+, \nll
T_{\pm} &=& \frac{1}{2}\,\left[ 1 + m_-^2 - m_+^2 \pm \lambda^{1/2}\lb 
1,m_-^2,m_+^2\rb \right],
\end{eqnarray}

\noindent
where we have introduced the corresponding K\"allen's $\lambda$-function,
whose positivity in turns implies

\begin{equation}
0 \leq  m_-^2 \leq  1, \qquad 0 \leq  m_+^2 \leq  \lb 1 - m_-\rb^2.
\end{equation}

\noindent
Similarly we obtain $-\ti + T'_- \leq  \tiii \leq  -\ti + T'_+$ with

\begin{equation}
T'_{\pm} = \frac{1}{2}\,\left[ e_1 + e_3 \pm \lambda^{1/2}\lb 1,m^2,
1+m^2-e_1-e_2\rb \right],
\end{equation}

\noindent
with the further constraint that

\begin{equation}
\lambda\lb 1,m^2,1+m^2-e_1-e_2\rb  = \lb e_1+e_3\rb^2 -
4\,m^2 \geq  0,
\end{equation}

\noindent
or $-\ti + T"_- \leq  \tiv \leq  -\ti + T"_+$ with

\begin{equation}
T"_{\pm} = \frac{1}{2}\,\left[ e_1 - e_4 \pm \lambda^{1/2}\lb 1,
e_1-m_-^2-m^2,e_3-m_+^2-m^2\rb \right],
\end{equation}

\noindent
with the additional condition that

\begin{equation}
\lambda\lb 1,e_1-m_-^2-m^2,e_3-m_+^2-m^2\rb  \geq  0
\end{equation}

\noindent
The other constraints will be described directly in the presence of 
kinematical cuts. First of all cuts on the final state invariant masses 
are very easy to implement thanks to our choice of integration variables,
eq.(\ref{nrps}). Next we introduce a threshold for all energies. Thus
if we require

\begin{equation}
E_i \geq  l_i\,\sqrt{s},
\end{equation}

\noindent
this will be the same as

\begin{equation}
\Delta\,t_i \leq  x_+e_i-2\,l_i.
\end{equation}

\noindent
Let us discuss explicitly the case of $\Delta$ positive, we get

\begin{eqnarray}
t_i &\leq & Xe_i-L_i, \nll
X &=& {{x_+}\over {\Delta}}, \qquad L_i = 2\,{{l_i}\over {\Delta}}.
\end{eqnarray}

\noindent
Requiring consistency among these four conditions we obtain

\begin{eqnarray}
1-\tw-Xe_4+L_4 &\leq & \tiii \leq  Xe_3-L_3, \nll
\tw-Xe_2+L_2 &\leq & \ti \leq  Xe_1-L_1, \nll
1-X(e_3+e_4)+L_3+L_4 &\leq & \tw \leq  X(e_1+e_2)-L_1-L_2, 
\end{eqnarray}

\noindent
with a derived condition on the radiated energy,

\begin{equation}
x_+-x_- \geq  \frac{1}{2}\lb 1 + \sum_i\,L_i\rb .
\label{rad}
\end{equation}

\noindent
Next let us require that $c_i \leq  \cos\theta_i \leq  C_i$, where $\theta_i$ 
is the i-th scattering angle. First we define

\begin{equation}
\gamma_i = \frac{1}{2}\lb 1 - c_i\rb , \qquad
\Gamma_i = \frac{1}{2}\lb 1 - C_i\rb .
\end{equation}

\noindent
In terms of invariants the constraints become

\begin{equation}
\sigma_i\,e_i \leq  t_i \leq  \Sigma_i\,e_i,
\end{equation}

\noindent
with the following definition

\begin{eqnarray}
{1\over {\sigma_i}} &=& 1 + {{\gamma_i}\over {1-\gamma_i}}\,
{{x_-}\over {x_+}}, \nll
{1\over {\sum_i}} &=& 1 + {{\Gamma_i}\over {1-\Gamma_i}}\,
{{x_-}\over {x_+}}.
\end{eqnarray}

\noindent
Eq. (\ref{rad})will translate into another set of constraints

\begin{eqnarray}
\tiii &\geq & \max\lb \sigma_3e_3,1-\Sigma_4e_4-\tw\rb , \nll
\tiii &\leq & \min\lb \Sigma_3e_3,1-\sigma_4e_4-\tw\rb , \nll
\ti &\geq & \max\lb \sigma_1e_1,\tw-\Sigma_2e_2\rb , \nll
\ti &\leq & \min\lb \Sigma_1,e_1,\tw-\sigma_2e_2\rb ,\nll
\tw &\geq & \max\lb 1-\Sigma_3e_3-\Sigma_4e_4,\sigma_1e_1+\sigma_2e_2\rb , \nll
\tw &\leq & \min\lb 1-\sigma_3e_3-\sigma_4e_4,\Sigma_1e_1+\Sigma_2e_2\rb.
\end{eqnarray}

\noindent
The optimal use of a deterministic integration routine (fixed grid) is 
largely based on the possibility of backward propagation of the
constraints. Typically we deal with an integral

\begin{equation}
\int\,dx_1\,\theta\lb b_1-x_1\rb \lb x_1-a_1\rb \dots \dots \int\,dx_n\,
\theta\left[b_n\lb x_1,\dots,x_{n-1}\rb - x_n\right]
\theta\left[x_n - a_n\lb x_1,\dots,x_{n-1}\rb\right].
\end{equation}

\noindent
Solving the constraint $b_n \geq a_n$ in terms of $x_{n-1}$ gives
$c_n(x_1,\dots,x_{n-2}) \leq x_{n-1} \leq d_n(x_1,\dots,x_{n-2})$, so that
we have now to solve

\begin{eqnarray}
a_{n-1}\lb x_1,\dots,x_{n-2}\rb &\leq& b_{n-1}\lb x_1,\dots,x_{n-2}\rb, \nll
c_{n}\lb x_1,\dots,x_{n-2}\rb &\leq& b_{n-1}\lb x_1,\dots,x_{n-2}\rb, \nll
a_{n-1}\lb x_1,\dots,x_{n-2}\rb &\leq& d_n\lb x_1,\dots,x_{n-2}\rb, \nll
c_n\lb x_1,\dots,x_{n-2}\rb &\leq& d_n\lb x_1,\dots,x_{n-2}\rb,
\end{eqnarray}

\noindent
and to iterate until the most external integration is reached. In our case
all the constraints, including the non-linear one, are at most quadratic
in the integration variables so that a complete solution of the
procedure is available. However the rapid growth in the number of the 
constraints, expecially when several kinematical cuts are imposed, and
the eventual occurrence of some splitting of the integrals ($[a \leq x_i \leq 
b]\cup [c \leq x_i \leq d]$) suggests an intermediate, hybrid, procedure where 
some of the constraints are left for a numerical control. Sometimes it is more 
convenient, in terms of computational speed, to increase the number of points 
in the fixed grid and to {\it lose} some of them instead of having a smaller 
number of points and full efficiency. This is also based on the fact that 
computing the amplitudes is usually faster than implementing the {\it next} 
level of backward propagation of the constraints.

\subsection{The Matrix Elements}

The generic process that we want to compute is characterized by six
external fermionic legs and by a considerable number of diagrams, up
to $64$ diagrams for final state identical particles not including
electrons or Higgs boson exchange. Already long time ago several 
authors~\cite{amp} realized that it is
extremely unrealistic to write down the full matrix element $M$,
to square it and to sum over polarizations. As a consequence of this fact
various groups have developed their own way of evaluating the helicity
amplitudes, each as a single complex number to be squared numerically.
Our approach is essentially based on the request that everything has
to be expressed in terms of the invariants which specify the process.
To briefly summarize the main features of the method we first define

\begin{eqnarray}
u_{\lambda}(p) &=& \Pi_{\lambda}\,u(p), \qquad
\baru_{\lambda}(p) = \baru(p)\,\Pi_{-\lambda}, \nll
v_{\lambda}(p) &=& \Pi_{-\lambda}\,v(p), \qquad
\barv_{\lambda}(p) = \barv(p)\,\Pi_{\lambda}, \nll
\Pi_{\lambda} &=& \frac{1}{2}\,\lb 1 + \lambda\gamma^5\rb , \quad 
\lambda = \pm 1,
\end{eqnarray}

\noindent
where $u,v$ are Dirac (massless) spinors. The key ingredient in evaluating
helicity amplitudes is given by the following set of formulas where all the
relevant operators are explicitly given.

\begin{eqnarray}
v_{\lambda}(p)\,\baru_{\sigma}(q) &=& - \lb 2\,p\cdot q\rb^{-1/2}\,
\Pi_{-\lambda}\,\left[\delta_{\lambda,\sigma}\psla\qsla +
\delta_{\lambda,-\sigma}\psla\nsla\qsla\right], \nll
u_{\lambda}(p)\,\baru_{\sigma}(q) &=& - \lb 2\,p\cdot q\rb^{-1/2}\,
\Pi_{\lambda}\,\left[\delta_{\lambda,\sigma}\psla\nsla\qsla +
\delta_{\lambda,-\sigma}\psla\qsla\right], \nll
v_{\lambda}(p)\,\barv_{\sigma}(q) &=& - \lb 2\,p\cdot q\rb^{-1/2}\,
\Pi_{-\lambda}\,\left[\delta_{\lambda,\sigma}\psla\nsla\qsla +
\delta_{\lambda,-\sigma}\psla\qsla\right], 
\end{eqnarray}

\noindent
where we have introduced an {\it auxiliary} vector $n_{\mu}$ which
satisfies $n\cdot n = 1$, $n\cdot p = n\cdot q = 0$ and it is otherwise
{\it arbitrary}. In the previous relation an overall phase has been
constantly neglected, i.e. these relations are exact modulus an arbitrary
phase, for instance the phase of $\baru_{\lambda}(p)u_{\lambda}(q)$. 
This fact alone will be of no consequence as long as one remembers
to organize the calculation of all the diagrams contributing to a certain
helicity amplitude (for a given process) in such a way that the unknown phase 
remains as an {\it overall} phase. Thus in WTO all the processes
$e^+(p_+)e^-(p_-) \to f_1(q_1) \barf_2(q_2) f_3(q_3) \barf_4(q_4)$ are 
organized in such a way that the requested operators are always relative
to the pairs $(p_-,q_3)$, $(p_+,q_2)$ and $(q_1,q_4)$. Following this strategy
and given that we still want a final answer completely expressible
in terms of invariants, we have made a very specific choice of the three
auxiliary vectors which we need. They are

\begin{eqnarray} 
l_{\mu} &=& {1\over {N_l}}\veps_{\mu\nu\alpha\beta}\,
q_3^{\nu}q_2^{\alpha}p_+^{\beta}, \nll
m_{\mu} &=& {1\over {N_m}}\veps_{\mu\nu\alpha\beta}\,
q_2^{\nu}q_4^{\alpha}q_1^{\beta}, \nll
n_{\mu} &=& {1\over {N_n}}\veps_{\mu\nu\alpha\beta}\,
q_2^{\nu}p_-^{\alpha}q_3^{\beta}, 
\end{eqnarray}

\noindent
where the proper normalization has been included, i.e.

\begin{eqnarray}
N_l &=& \frac{1}{2}\,\lb x_{_{14}}x_{_{15}}x_{_{45}}\rb^{1/2},  \nll
N_m &=& \frac{1}{2}\,\lb x_{_{34}}x_{_{36}}x_{_{46}}\rb^{1/2},  \nll
N_n &=& \frac{1}{2}\,\lb x_{_{24}}x_{_{25}}x_{_{45}}\rb^{1/2}.
\end{eqnarray}

\noindent
Actually only certain combinations of the auxiliary vectors appear
explicitly in the calculation. For instance $\psla_-\nsla\qsla_3$ etc. 
We make use of the relation

\begin{equation}
\gamma^{\mu}\veps_{\mu\nu\alpha\beta} = \gamma_{\nu}\gamma_{\alpha}
\gamma_{\beta}\gamma^5 + \gamma^5\lb \delta_{\nu\alpha}\gamma_{\beta}
- \delta_{\nu\beta}\gamma_{\alpha} + \delta_{\alpha\beta}\gamma_{\nu}\rb ,
\end{equation}

\noindent
to obtain

\begin{eqnarray}
\qsla_2\lsla\psla_+ &=& \lb  -{{x_{_{14}}}\over {x_{_{15}}x_{_{45}}}}\rb^{1/2}\,
\gamma^5\qsla_2\qsla_3\psla_+,  \nll
\qsla_4\msla\qsla_1 &=& \lb  -{{x_{_{36}}}\over {x_{_{34}}x_{_{46}}}}\rb^{1/2}\,
\gamma^5\qsla_4\qsla_2\qsla_1,  \nll
\psla_-\nsla\qsla_3 &=& \lb  -{{x_{_{25}}}\over {x_{_{24}}x_{_{45}}}}\rb^{1/2}\,
\gamma^5\psla_-\qsla_2\qsla_3.
\end{eqnarray}

\noindent
where again all particles are strictly massless. It is relatively easy
to show that no singularity will be introduced by the normalization factors
at the boundaries of the phase space. However from a numerical point of
view strong cancellation may occur for particular processes, expecially
when both $l$ and $n$ will appear giving a factor $x_{45}^{-1}$ in the
amplitude. To avoid numerical instabilities we have made a systematic
use of the freedom in choosing the auxiliary vectors, namely

\begin{eqnarray} 
l_{\mu} &=& {1\over {N_l}}\veps_{\mu\nu\alpha\beta}\,
k_a^{\nu}q_2^{\alpha}p_+^{\beta}, \nll
m_{\mu} &=& {1\over {N_m}}\veps_{\mu\nu\alpha\beta}\,
k_b^{\nu}q_4^{\alpha}q_1^{\beta}, \nll
n_{\mu} &=& {1\over {N_n}}\veps_{\mu\nu\alpha\beta}\,
k_c^{\nu}p_-^{\alpha}q_3^{\beta}, 
\end{eqnarray}

\noindent
where $k_a$ is any linear combination of $p_-,q_1,q_3,q_4\,$, $k_b$ of
$p_+,p_-,q_2,q_3$ and $k_c$ of $p_+,q_1,q_2,q_4$. Whenever a region of
the phase-space is examined, which is dangerously closed to some
portion of the boundary, then we have switched to a choice of the
auxiliary vectors free of numerical instabilities in that region.
Let us discuss this point in more details since it is connected with
another important issue. Whenever we change auxiliary vectors for a
particular diagram we must also remember that our amplitudes are defined
modulus a phase, thus to implement this procedure we have to compute
the relative phase. For instance we redefine

\begin{eqnarray} 
L_{\mu} &=& {1\over {N_L}}\veps_{\mu\nu\alpha\beta}\,
p_-^{\nu}q_2^{\alpha}p_+^{\beta}, \nll
M_{\mu} &=& {1\over {N_M}}\veps_{\mu\nu\alpha\beta}\,
q_3^{\nu}q_4^{\alpha}q_1^{\beta}, \nll
N_{\mu} &=& {1\over {N_N}}\veps_{\mu\nu\alpha\beta}\,
p_+^{\nu}p_-^{\alpha}q_3^{\beta}, 
\end{eqnarray}

\noindent
and assume that for a given diagram we need to switch from

\begin{eqnarray}
\barv_{\lambda}(q_2)\lsla v_{\lambda}(p_+) \quad &\to& \quad
\barv_{\lambda}(q_2)\clsla v_{\lambda}(p_+),  \nll
\barv_{\lambda}(q_4)\msla u_{-\lambda}(q_1) \quad &\to& \quad
\barv_{\lambda}(q_4)\cmsla u_{-\lambda}(q_1),  \nll
\baru_{-\lambda}(p_-)\nsla u_{-\lambda}(q_3) \quad &\to& \quad
\baru_{-\lambda}(p_-)\cnsla u_{-\lambda}(q_3).
\end{eqnarray}

\noindent
First of all we introduce

\begin{eqnarray}
\barv_{\lambda}(q_2)\lsla v_{\lambda}(p_+) &=& e^{i\,\phi_l}\,X_l,  \nll
\barv_{\lambda}(q_2)\clsla v_{\lambda}(p_+) &=& e^{i\,\phi_L}\,X_L,  
\end{eqnarray}

\noindent
and we compute

\begin{eqnarray}
e^{i(\phi_l-\phi_L)}\,X_lX_L &=& tr\lb \lsla\psla_+\clsla\qsla_2\rb , \nll
X_l^2(X_L^2) &=& tr\left[\lsla(\clsla)\psla_+\clsla(\lsla)\qsla_2\rb .
\end{eqnarray}

\noindent
In this context it is easy to show that

\begin{eqnarray}
e^{i(\phi_l-\phi_L)} &=& e^{-i(\phi_n-\phi_N)} = 
\frac{1}{2}\,{1\over {\lb x_{_{15}}x_{_{24}}x_{_{45}}\rb^{1/2}}}\,
\lb x_{_{15}}x_{24}+x_{_{45}}-x_{_{14}}x_{_{25}} - 4\,i\lambda\veps_4\rb , \nll
e^{i(\phi_m-\phi_M)} &=&
\frac{1}{2}\,{1\over {\lb x_{_{34}}x_{_{35}}x_{_{46}}x_{_{56}}\rb^{1/2}}}\,
\lb x_{_{34}}x_{_{56}}+x_{_{35}}x_{_{46}}-x_{_{36}}x_{_{45}} + 4\,i
\lambda\veps_{15}\rb , \nll
\veps_{15} &=& \veps\lb q_1,q_2,q_3,q_4\rb . \nll
\end{eqnarray}

\noindent
A typical example of a diagram to evaluated will be

\begin{equation}
D(\lambda,\rho,\sigma) = \barv_{\lambda}(p_+)\Gamma u_{-\lambda}(p_-)
\baru_{\rho}(q_1)\Gamma' v_{-\rho}(q_2)\baru_{\xi}(q_3)\Gamma"
v_{-\xi}(q_4),
\end{equation}

\noindent
where $\Gamma,\dots\Gamma"$ are strings of $\gamma-$functions. This will be 
computed as

\begin{eqnarray}
D(\lambda,\rho,\sigma) &=& tr\left\{\Gamma'\left[v_{-\rho}(q_2)\barv_{\lambda}
(p_+)\right]\Gamma\left[u_{-\lambda}(p_-)\baru_{\xi}(q_3)\right]
\Gamma"\left[v_{-\xi}(q_4)\baru_{\rho}(q_1)\right]\right\}  \nll
{}&=& \sum_{\lambda,\rho,\xi=-1}^{+1}\left[A_{\lambda,\rho,\xi} +
i\,\sum_{i=1,5}\,b^i_{\lambda,\rho,\xi}\,\veps_i\right].
\end{eqnarray}

\noindent
The presence of saturated $\veps$-tensors seems to represent an obstacle
towards giving the full amplitude in terms of invariants. Our strategy
therefore is to express all $|\veps|$ in terms of invariants. For instance
if we define $\veps_1 = \veps(p_+,p_-,q_1,q_2)$ then we obtain

\begin{eqnarray}
|\veps|^2 &=& 2\lb x_{_{13}}x_{_{14}}x_{_{23}}x_{_{24}}+
x_{_{13}}x_{_{24}}x_{_{34}}+x_{_{14}}x_{_{23}}x_{_{34}}\rb  \nll
 & & \mbox{} - x_{_{13}}^2x_{_{24}}^2-x_{_{14}}^2x_{_{23}}^2-x_{_{34}}^2,
\end{eqnarray}

\noindent
and similar expressions for the other $5$ independent saturated $\veps$-tensors.

\begin{eqnarray}
\veps_1 &=& \veps(p_+,p_-,q_1,q_2), \nll
\veps_2 &=& \veps(p_+,p_-,q_1,q_3), \nll
\veps_3 &=& \veps(p_+,p_-,q_2,q_3), \nll
\veps_4 &=& \veps(p_+,q_1,q_2,q_3), \nll
\veps_5 &=& \veps(p_-,q_1,q_2,q_3).
\end{eqnarray}

\noindent
Moreover the sign of $\veps_1\veps_2,\dots,\veps_1\veps_5$ can all be 
determined in terms of invariants. Let $\eta$ be the unknown sign of
$\veps_1$, then we end up with some expression of the form

\begin{equation}
d(\lambda,\rho,\xi) = \left[A_{\lambda,\rho,\xi} + i\,\eta\sum_{i=1,5}\,
b^i_{\lambda,\rho,\xi}\,{\hbox{sign}}\lb \veps_1\veps_i\rb |\veps_i|
\right].
\end{equation}

\noindent
Therefore, as long as we only need $|d|^2$, there will be no problem at all
with the unknown $\eta$ since $\eta^2= 1$.
Of course we can have additional imaginary parts in the amplitude, due to
the width of the vector bosons. Suppose that in a diagram some internal line
can be either a photon or a $Z$. Schematically we denote with
$a_{_V} + ib_{_V}\veps, V=\gamma,Z$ the corresponding amplitude and
with $\alpha + i\beta$ the $Z$ propagator. The overall expression for the
amplitude is therefore

\begin{equation}
d = \lb a_{\gamma}+\alpha a_{_Z}\rb - \beta b_{_Z}\veps + i\,\lb b_{\gamma} +
\beta a_{_Z}\rb\veps + i\,\beta a_{_Z},
\end{equation}

\noindent
In computing $|d|^2$ all terms which are linear in $\veps$ do not contribute
after integration, therefore

\begin{equation}
|d|^2 =  \lb a_{\gamma}+\alpha a_{_Z}\rb^2 + \lb\beta b_{_Z}\veps\rb^2 + 
\lb b_{\gamma} + \beta a_{_Z}\rb^2\veps^2 + \lb\beta a_{_Z}\rb^2.
\end{equation}

\noindent
We give an example of how the helicity amplitudes are worked out explicitly. 
Let us start from some diagram whose expression is

\begin{eqnarray}
\omega\lb \lambda,\rho,\xi\rb  &=& \barv_{-\lambda}(p_+)\gamma^{\mu}
\lb a+b\,\gamma^5\rb u_{\lambda}(p_-)  \nll
 & & \mbox{} \times \baru_{\rho}(q_3)\gamma^{\nu}\gamma_+\lb \psla_++\psla_--
\qsla_4\rb \gamma^{\mu}\lb a+b\,\gamma^5\rb v_{-\rho}(q_4)  \nll
 & & \mbox{} \times \baru_{\xi}(q_1)\gamma^{nu}\gamma_+v_{-\xi}(q_2),
\end{eqnarray}

\noindent
where $\gamma_{\pm} = 1 \pm \gamma^5$. It is easy to show that

\begin{eqnarray}
\omega\lb \lambda,\lambda,\lambda\rb  &=& 16\,\gl^2\delta_{\lambda,1} 
\lb x_{15}x_{24}x_{34}x_{45}x_{46}\rb^{-1/2} \nll
 & & \mbox{}\times tr\left[\Pi_{-\lambda}\gamma^{\mu}\gamma^5\psla_-\qsla_2
\qsla_3
\gamma^{\nu}\lb \psla_++\psla_--\qsla_4\rb \gamma^{\mu}\gamma^5
\qsla_4\qsla_2\qsla_1\gamma^{\nu}\gamma^5\qsla_2\qsla_3\psla_+\right],  \nll
{}&=& 16\,\gl^2\delta_{\lambda,1} 
\lb x_{15}x_{24}x_{34}x_{45}x_{46}\rb^{-1/2} \nll
 & & \mbox{} \times tr\left[\Pi_{-\lambda}\gamma^{\mu}\psla_-\qsla_2\qsla_3
\gamma^{\nu}\lb \psla_++\psla_--\qsla_4\rb \gamma^{\mu}
\qsla_4\qsla_2\qsla_1\gamma^{\nu}\qsla_2\qsla_3\psla_+\right],  \nll
\omega\lb \lambda,\lambda,-\lambda\rb  &=& 
\omega\lb \lambda,-\lambda,\lambda\rb  = 0,  \nll
\omega\lb \lambda,-\lambda,-\lambda\rb  &=& 16\,\gl\gr\delta_{\lambda,-1}
\lb x_{14}x_{25}x_{34}x_{46}\rb^{-1/2} \nll
 & & \mbox{}\times tr\left[\Pi_{-\lambda}\gamma^{\mu}\psla_-\qsla_3\gamma^{\nu}
\lb \psla_++\psla_--\qsla_4\rb \gamma^{\mu}\gamma^5\qsla_4\qsla_2\qsla_1
\gamma^{\nu}\qsla_2\psla_+\right] \nll
{}&=& 16\,\gl\gr\delta_{\lambda,-1}
\lb x_{14}x_{25}x_{34}x_{46}\rb^{-1/2} \nll
 & & \mbox{}\times tr\left[\Pi_{-\lambda}\gamma^{\mu}\psla_-\qsla_3\gamma^{\nu}
\lb \psla_++\psla_--\qsla_4\rb \gamma^{\mu}\qsla_4\qsla_2\qsla_1
\gamma^{\nu}\qsla_2\psla_+\right],
\end{eqnarray}

\noindent
where

\begin{equation}
\gl = \frac{1}{2}\lb a+b\rb , \qquad \gr = \frac{1}{2}\lb a-b\rb ,
\end{equation}

\noindent
after which the Kahane identities can be applied and the trace can be taken.
For instance we find

\begin{eqnarray}
\omega\lb \lambda,-\lambda,-\lambda\rb  &=& -32\,\gl\gr
\delta_{\lambda,-1}\lb x_{14}x_{25}x_{34}x_{46}\rb^{-1/2} \nll
 & & \mbox{}\times tr\left[\Pi_{-\lambda}\gamma^{\mu}\psla_-\qsla_3\qsla_1
\qsla_2
\qsla_4\gamma^{\mu}\lb \psla_++\psla_--\qsla_4\rb \qsla_2\psla_+\right] \nll
{}&=&  64\,\gl\gr
\delta_{\lambda,-1}\lb x_{14}x_{25}x_{34}x_{46}\rb^{-1/2} \nll
 & & \mbox{}\times tr\left[\Pi_{-\lambda}\qsla_4\qsla_2\qsla_1\qsla_3\psla_-
\lb \psla_++\psla_--\qsla_4\rb \qsla_2\psla_+\right].
\end{eqnarray}

\noindent
There is another place where we must worry about relative phases. Consider
the NC64 processes, typically $e^+e^- \to f\barf f \barf$. The $64$
diagrams arise from $4$ permutations of $16$ different topologies, which
we denote by

\begin{eqnarray}
{\cal P}_1\{q_1,q_2,q_3,q_4\} &=& \{q_1,q_2,q_3,q_4\},  \nll
{\cal P}_2\{q_1,q_2,q_3,q_4\} &=& \{q_3,q_2,q_1,q_4\},  \nll
{\cal P}_3\{q_1,q_2,q_3,q_4\} &=& \{q_1,q_4,q_3,q_2\},  \nll
{\cal P}_1\{q_1,q_2,q_3,q_4\} &=& \{q_3,q_4,q_1,q_2\}. 
\end{eqnarray}

\noindent
In the massless case there are $12$ distinct sets of helicity states which
we denote by labels $h = a-m$. For instance the effect of ${\cal P}_2$ can
be inferred from table $1$.

\begin{table}[hbtp]
\begin{center}
\begin{tabular}{|c|c|c|c|c|c|c|c|c|c|c|c|c|c|c|}
\hline
$p_+$ & $p_-$ & $q_1$ & $q_2$ & $q_3$ & $q_4$ &h& ${\cal P}_2\to$& 
$p_+$ & $p_-$ & $q_1$ & $q_2$ & $q_3$ & $q_4$ &h \\
$+$& $-$& $+$& $-$& $+$& $-$&  a& & $+$& $-$& $+$& $-$& $+$& $-$& a \\
$-$& $+$& $-$& $+$& $-$& $+$&  b& & $-$& $+$& $-$& $+$& $-$& $+$& b \\
$+$& $-$& $-$& $+$& $-$& $+$&  c& & $+$& $-$& $-$& $+$& $-$& $+$& c \\
$-$& $+$& $+$& $-$& $+$& $-$&  d& & $-$& $+$& $+$& $-$& $+$& $-$& d \\
$+$& $-$& $+$& $-$& $-$& $+$&  e& & $+$& $-$& $-$& $-$& $+$& $+$& i \\
$-$& $+$& $-$& $+$& $+$& $-$&  f& & $-$& $+$& $+$& $+$& $-$& $-$& j \\
$+$& $-$& $-$& $+$& $+$& $-$&  g& & $+$& $-$& $+$& $+$& $-$& $-$& l \\
$-$& $+$& $+$& $-$& $-$& $+$&  h& & $-$& $+$& $-$& $-$& $+$& $+$& m \\
\hline
\hline
\end{tabular}
\caption{\it The effect of the ${\cal P}_2$ permutation on the non-zero 
helicity sets for the $16$ fundamental topologies contributing to a typical 
NC64 process.}
\end{center}
\label{tab1}
\end{table}

\noindent
Our strategy has been to compute explicitly the $16$ diagrams
corresponding to the standard ordering $\{q_1,q_2,q_3,q_4\}$ and
to perform the relative permutations of invariants and helicity sets
for the remaining diagrams. This however requires the knowledge of all
the relative phases which we have computed explicitly. Since the relative
phases depend on the particular assignment of momenta and helicities but
not on the particular topology under considerations it is enough to
compute them for one diagram per permutation.

As we have already stressed all fermions in WTO are kept rigorously
massless. This fact however should not be confused with a limitation
of the helicity formalism. Indeed we briefly sketch how fermion
masses can be incorporated into our approach. We consider spinors
$u(p,n,\lambda),\,\lambda = \pm 1$ defined to be the eigenstates
of the operator $P_+(p,n,\lambda)$ corresponding to eigenvalue $1$.

\begin{eqnarray}
P_+\lb p,n,\lambda\rb  &=& \Lambda_+(p)\frac{1}{2}\lb 1+i\lambda\gamma^5
\nsla\rb \Lambda_+(p),  \nll
\Lambda_+(p) &=& {1\over {2\,p_{_0}}}\,\lb -i\,\psla+m\rb \gamma^4, 
\end{eqnarray}

\noindent
where $n$ is the polarization vector and $p\cdot n = 0$, $n^2 = 1$. If we
consider the typical CC20 process, $e^+e^- \to e^- \barnu _e \nu_{\mu} 
\mu^+$ and we are interested in the finite $m_e$ effects in the limit
of zero scattering angle then things can be organized in such a way that
we only need to compute

\begin{equation}
u\lb p,n,\sigma\rb \baru\lb q,\lambda\rb , \qquad {\hbox{etc.}},
\end{equation}

\noindent
where $p^2 = - m^2$ while $q^2 = 0$. In this case we easily obtain

\begin{eqnarray}
u\lb p,n,\sigma\rb \baru\lb q,\lambda\rb  &=& \frac{1}{8}\lb 
mp_{_0}q_{_0}\rb^{-1/2}  \nll
 & & \mbox{}\times \sum_{a=\pm}\,N_a(p,q)\lb -i\,\psla+m\rb 
\lb 1+ia\,\nsla\rb \qsla\lb 1+\lambda\gamma^5\rb 
{\cal P}^a(\lambda,\sigma),  \nll
m\,N_a^{-2} &=& am\,q\cdot n - p\cdot k, \qquad {\cal P}^a(\lambda,\sigma) =
\frac{1}{2}\lb 1+a\lambda\sigma\rb .
\end{eqnarray}

\noindent
Moreover if both fermionic lines are massive we can repeat the whole
procedure by using

\begin{eqnarray}
u\lb p,n,-\lambda\rb \baru\lb q,n,\lambda\rb &=& {1\over N^2}\,
\lb -i\psla+m\rb\lb 1-i\lambda\gamma^5\nsla\rb\lb -i\qsla+m\rb
\lb 1+i\lambda\gamma^5\nsla\rb,  \nll
u\lb p,n,\lambda\rb \baru\lb q,n,\lambda\rb &=& {1\over N^2}\,
\lb -i\psla+m\rb\lb 1+i\lambda\gamma^5\nsla\rb\nsla\lb -i\qsla+m\rb
\lb 1+i\lambda\gamma^5\nsla\rb,  
\end{eqnarray}

\noindent
with
\begin{equation}
N^2 = \lb p+q\rb^2.
\end{equation}

\noindent
For NC32 and non-leptonic NC64 or Mix43 the exchange of gluons must be
taken into account. This raises the basic question of what to use for
$\alpha_s$ or, alternatively, at which scale to compute it. Here the
situation is far from satisfactory, expecially as compared to the well
established predictions for $2$-fermion production at the $Z$ resonance.
In the present version of WTO we have adopted the {\it ad hoc} prescription
of using $\alpha_s(\wm)$ which becomes therefore an input parameter. 
However a more satisfactory choice would imply to consider a {\it running}
$\alpha_s$. Ideally for each sub-process $g^*(s) \to {\bar q}q$ we should
include a factor $\alpha_s(s)$ and subsequently integrate over the invariant 
mass $s$, however this can only be done when a reasonable cut is applied,
$s \geq s_{_0}$ to avoid entering non-perturbative regions where the 
correction factor would go out of control. Clearly more refined analyses are 
needed to reach a satisfactory level of description.
We end this section with a very short description of what it is meant by NQCD.
Consider the CC11 process $e^+e^- \to \mu^-\barnu_{\mu}u\bard$ which
effectively is a CC10 process. One would like to include final state QCD
corrections, even when kinematical cuts are imposed. By {\it naive} QCD
we mean a simple recipe where the total $W$-width is corrected by a factor

\begin{equation}
\Gamma_{_W} \to \Gamma_{_W}\,\lb 1 + \frac{2}{3}\,{{\alpha_s(\wm)}
\over {\pi}}\rb,
\end{equation}

\noindent
and where the cross section gets multiplied by

\begin{equation}
\sigma_{CC10,C} \to \sigma_{CC10,C}\,\lb 1 + {{\alpha_s(\wm)}
\over {\pi}}\rb.
\end{equation}

\noindent
This {\it naive} approach, consequence of our ignorance about the complete
result, would be correct only for $\sigma_{CC03,ex}$, the double-resonant
approximation with fully extrapolated setup. For $\sigma_{CC10,C}$ it
is instead only a rough approximation because of two reasons. First of all in
CC10 we have not only a virtual QCD correction to the $Wu\bard$ vertex but 
also a box diagram. Moreover QED and QCD radiation are quite different if
cuts are imposed, expecially in presence of severe cuts. Thus any inclusion
of final state QCD corrections is, at present, only a very crude approximation
which moreover can become quite bad whenever stringent kinematical cuts are 
applied to the process.

\subsection{Mappings of the phase space}

For a total cross section we need a $7$-dimensional ($9$-dimensional when
initial state radiation is included) integration. In order to achieve the
requested precision we must examine the structure of the integrand which will
show a complicated peaking structure due to propagators (both in the time-like
and in the space-like region), to Jacobians and to square-integrability but
not integrability of the structure functions. As a consequence we must
perform a careful analysis and find out, as much as possible, those 
isomorphisms of the phase space which are needed to cure the peaks of the 
differential distribution. First of all by
transforming the integration over the phase space into an integration over
the kinematically independent invariants we have introduced a Jacobian
which contains two inverse square roots of polynomials. This will always
be cured as follows. The part of the Jacobian denoted by $R_{_2}$ is a 
quadratic form in $\ti$ so that our cross section can be schematically written
as

\begin{equation}
\sigma = \int\,dm_-^2\dots d\tw\,\int_{t_l}^{t_u}\,d\ti\,
f(\dots,\ti)\lb -a\ti^2+2\,b\ti+c\rb^{-1/2},
\end{equation}

\noindent
where $a,b,c$ are functions of the other invariants and moreover $a>0$
over all the phase space. We introduce the following mapping

\begin{equation}
\ti \to t'_1 = -{1\over {\sqrt{a}}}\,\arccos{{b-a\ti}\over D}, \qquad
D^2 = b^2+ac.
\end{equation}

\noindent
which gives

\begin{eqnarray}
\sigma &=& \int\,dm_-^2\dots d\tw\,\int_{t'_l}^{t'_u}\,dt'_1\, 
f\lb \dots,b/a+D/a\sin(\sqrt{a}t'_1)\rb , \nll
t'_l &=& -{1\over {\sqrt{a}}}\,\arcsin\left[1 + {a\over D}
\lb t_+-t_l\rb \right], \nll
t'_u &=& {1\over {\sqrt{a}}}\,\arcsin\left[1 + {a\over D}
\lb t_u-t_-\rb \right], \nll
t_{\pm} &=& {b\over a} \mp {D\over a}.
\end{eqnarray}

\noindent
There is another square root in the Jacobian which is slightly more difficult
to treat since it is a polynomial of order $> 2$ in all invariants. Actually
it can be seen as a quartic form in the variable $z=m^2$ defined in
eq.(\ref{inv}), i.e.

\begin{equation}
R_4^2 = -\left[z^2 + \lb \Sigma - 1\rb z + \lambda_+\right]
\left[z^2 + \lb \Sigma - 1\rb z + \lambda_-\right],
\end{equation}

\noindent
where we have introduced

\begin{eqnarray}
\Sigma &=& m_-^2+m_+^2+M_0^2+m_0^2,  \nll
\lambda_{\pm} &=& M_0m_0 \pm m_-m_+.
\end{eqnarray}

\noindent
Given

\begin{eqnarray}
r_{1,2} &=& \frac{1}{2}\lb 1 - \Sigma \mp \sqrt{\Delta_-}\rb , \nll
s_{1,2} &=& \frac{1}{2}\lb 1 - \Sigma \mp \sqrt{\Delta_+}\rb , \nll
\Delta_{\pm} &=& \lb 1 - \Sigma\rb^2 - 4\,\lambda_{\pm}^2,
\end{eqnarray}

\noindent
we have the following three possibilities

\begin{equation}
\Delta_- \leq 0, \qquad  \Delta_- \geq 0 \geq \Delta_+, \qquad
\Delta_+ \geq 0, 
\end{equation}

\noindent
which correspond to

\begin{eqnarray}
R_4^2 &\leq& 0, \nll
r_1 &\leq& z \leq r_2, \nll
r_1 &\leq& z \leq s_1, \quad {\hbox{or}} \quad
s_2 \leq z \leq r_2.
\end{eqnarray}

\noindent
Only the second one will be discussed here in some details. Again 
schematically we have

\begin{eqnarray}
I &=& \int_{r_l}^{r_u}\,dz {{f(z,\dots)}\over {R_4}},  \nll
r_l &\geq& r_1, \qquad r_u \leq r_2, \qquad r_{1,2} = r \mp \frac{1}{2}
\sqrt{\Delta_-}.
\end{eqnarray}

\noindent
Next we introduce a new variable

\begin{equation}
\cos\phi = 2\,{{r-z}\over {\sqrt{\Delta_-}}},
\end{equation}

\noindent
and distinguish among three alternatives. If $r_u \leq r$ then

\begin{equation}
I = \int_{\mu_l}^{\mu_u}\,d\mu\,f\lb r-\frac{1}{2}\sqrt{\Delta_-}\cos\phi,
\dots\rb , \qquad \mu = \frac{\delta}{\gamma}\,F(\phi,K).
\end{equation}

\noindent
If instead $r_l \geq r$ then

\begin{equation}
I = \int_{\mu_u}^{\mu_l}\,d\mu\,f\lb r+\frac{1}{2}\sqrt{\Delta_-}\cos\phi,
\dots\rb , \qquad \mu = \frac{\delta}{\gamma}\,F(\phi,K).
\end{equation}

\noindent
Finally if $r_l \leq r \leq r_u$ we first introduce 

\begin{equation}
{\bar\mu} = - {{\delta}\over {\gamma}}\,F\lb \frac{\pi}{2},K\rb ,
\end{equation}

\noindent
and subsequently we obtain

\begin{equation}
I = \int_{\mu_l}^{\bar\mu}\,d\mu\,f\lb r-\frac{1}{2}\sqrt{\Delta_-}\cos\phi,
\dots\rb  + \int_{\mu_u}^{\bar\mu}\,d\mu\,f\lb r+\frac{1}{2}
\sqrt{\Delta_-}\cos\phi,\dots\rb .
\end{equation}

\noindent
In the last equations $F$ denotes the elliptic function 

\begin{equation}
F\lb \phi,K\rb = \int_0^{\phi}\,d\psi \lb 1 - K^2\sin^2\psi\rb^{-1/2},
\end{equation}

\noindent
and

\begin{eqnarray}
R_4^2 &=& -(z-r_1)(z-r_2)\left[(z-r)^2+s^2\right],  \nll
A &=& {{s^2+(r_2-r)(r_1-r)}\over {s\sqrt{\Delta_-}}},  \nll
K_1 &=& A + \lb A^2+1\rb^{1/2},  \nll
K^2 &=& {1\over {1+K_1^2}},  \nll
{{\delta}\over {\gamma}} &=& -2\,{{KK_1}\over {\lb 2\,sK_1\sqrt{\Delta_-}
\rb^{1/2}}}.
\end{eqnarray}

\noindent
The case of four real roots will not be considered explicitly here but
it can be treated along the same lines, i.e. by introducing elliptic
functions.

Whenever initial state radiation is included we have to consider the
following two integrations

\begin{equation}
\sigma = \int\,dx_+dx_-\,D(x_+,s),D(x_-,s)\,\int\,dPS\,\Theta_{cut} \,f,
\end{equation}

\noindent
where the structure functions are

\begin{equation}
D(x,s) = \frac{\beta}{2}\lb 1-x\rb^{\beta/2-1}\left\{ 
{{\exp\left[\frac{1}{2}\beta\lb \frac{3}{4}-\gamma_{_E}\rb\right]}\over
{\Gamma\lb 1+ \frac{1}{2}\beta\rb}} + D_r(x,s)\right\}.
\label{sf}
\end{equation}

\noindent
and the $\beta$-factor is 

\begin{equation}
\beta = 2\,{{\alpha}\over {\pi}}\lb \ln{s\over {m_e^2}} - 1\rb .
\end{equation}

\noindent
In order to have a proper treatment of the $x_+/x_-$ integration we first
change variables to 

\begin{equation}
x_- = u, \qquad x_+ = {v\over u}.
\end{equation}

\noindent
The behavior near $x = 1$ is cured by introducing two new variables, 
$ 0 \leq U,V \leq 1$ which are related to $u,v$ through

\begin{eqnarray}
u &=& 1 - \lb 1 - u_l\rb \lb 1 - U\rb^{2/\beta}, \nll
v &=& u - \lb u - u_l\rb \lb 1 - V\rb^{2/\beta},
\end{eqnarray}

\noindent
where we assume that the final state invariant masses are bounded by below,
i.e. $m_-^2 \geq M_1^2,\dots,m_+^2 \geq M_6^2$ and

\begin{eqnarray}
u_l &=& \max(u_a,u_b,u_c),  \nll
u_a &=& (M_1+M_6)^2,  \nll
u_b &=& (M_2+M_5)^2,  \nll
u_c &=& (M_3+M_4)^2.
\end{eqnarray}

\noindent
Finally most of the diagrams belonging to a given process receive
resonating contributions by vector boson propagators. The {\it resonating}
mapping is not always mandatory, expecially for those processes where
the final answer is not dominated by $W$ or $Z$ peaks, like for instance in
NC processes where we can have large effects from photon exchange or from
$t$-channel diagrams. However for CC processes we have a double-resonating
$W$-exchange which will show up as

\begin{eqnarray}
\sigma &=& \int\,\dots\int_{a_-}^{b_-}\,dm_-^2\int_{a_+}^{b_+}\,dm_+^2\,
{1\over {\left[\lb vm_-^2-\muw^2\rb^2+\lb vm_-\sgw\rb^2\right]
\left[\lb vm_+^2-\muw^2\rb^2+\lb vm_+\sgw\rb^2\right]}} \nll
 & & \mbox{}\times\int\,\dots f\lb m_-^2,m_+^2\dots\rb ,
\end{eqnarray}

\noindent
where $\muw^2= \wm^2/s$ and $\sgw = \gw/\wm$.
The lower and upper limits of integrations for $m_{\pm}^2$ may differ
from the natural limits because of kinematical cuts. Let us deal with one
specific integration and change variable, e.g. $z = vm_-^2$. The
typical integral will be

\begin{equation}
I = {1\over v}\,\int_{z_-}^{z_+}\,dz\,{{f(z,\dots)}\over
{(z-\muw^2)^2+(z\sgw)^2}}.
\end{equation}

\noindent
We first transform

\begin{equation}
z = {{\muw^2}\over {1+\lb {{\gw}\over {\wm}}\rb^2}}\,
\left\{1 + {{\gw}\over {\wm}}\,\tan\left[\lb z_b-z_a\rb Z+z_a\right]
\right\},
\end{equation}

\noindent
with

\begin{equation}
z_{a,b} = \arctan{{z_{\mp}-\muw^2 + \lb {{\gw}\over {\wm}}\rb^2
z_{\mp}}\over {\sgw\muw}},
\end{equation}

\noindent
and where $0 \leq Z \leq 1$. In this way we get

\begin{equation}
I = {{z_b-z_a}\over {\sgw\muw}}\,\int_0^1\,dZ\,f\lb z(Z),\dots\rb .
\end{equation}

\noindent
Concerning the resonating mapping we stress that indeed there are $4$-fermion
processes which receive the largest contribution from double-resonating
diagrams. A typical example is given by $e^+e^- \to \mu^-\barnu _{\mu}u
{\bar d}d$ at LEP 2 energies where the percentage difference between
CC03 and CC11 is totally negligible. In this and similar situations the
isomorphism of the phase space that we have described is mandatory.
The other cases where the mapping is strongly advised are those where very 
stringent cuts are applied around the double resonance. For instance

\begin{equation}
e^+e^- \to {\bar b}b{\bar f}f, \qquad \zm-\Delta \leq M(\barf f) \leq
\zm+\Delta, \quad M({\bar b}b) \geq M_{_0}.
\end{equation}

\noindent
On the contrary there are examples where double-resonating diagrams only
give a small fraction of the total answer, either because there are
$\gamma$-exchanges not suppressed by kinematical cuts or because of
large $t$-{\it channel} contributions. In all these cases it turns out to be
more convenient to deal with a {\it flat} phase-space, i.e. no mapping at
all.

\subsection{Gauge invariance}

Any calculation for $e^+e^- \to 4$-fermions is only nominally a {\it tree
level} approximation because of the presence of charged and neutral, unstable 
vector bosons and of their interaction with photons.
Unstable particles require
a special care and their propagators, in some channels, must necessarily
include an imaginary part or in other words the corresponding $S$-matrix
elements will show poles shifted into the complex plane. In any 
field-theoretical approach these imaginary parts are obtained by performing
the proper Dyson resummation of the relative two-point functions, which
at certain thresholds will develop the requested imaginary component.
It is well known that the bosonic parts of vector boson self-energies are
not separately gauge independent, or stated differently they will show a
gauge-parameter dependence. This problem will arise in the context of a full
one loop calculation, while here we can take into account only the fermionic
parts which are obviously free of ambiguities. Thus the correct recipe
seems representable by a Dyson resummation of fermionic self-energies where
only the imaginary parts are actually included. As a result the vector boson
propagators will be inserted into the corresponding tree level amplitudes with
a $p^2$-dependent width. Its has already been noticed by several
authors that even this simple idea gives rise to a series of inconsistencies,
which sometimes may give results completely inconsistent even from a numerical
point. The fact is that the introduction of a width into the propagators will
inevitably result, in some cases, into a breakdown of the relevant Ward 
identities of the theory with a consequent violation of some well understood
cancellation mechanism. Sometimes the effect of spoiling a cancellation 
among diagrams can result into a numerical catastrophe.

In the present version WTO is including reparation of gauge invariance
according to the {\it fermion-loop} scheme~\cite{gi} for the CC20 family,
while an extension to other processes in the TeV region is currently
in preparation.

Here we briefly sketch the procedure adopted for CC20. In this case there
is a violation of gauge invariance induced by the so called {\it fusion}
diagram which contributes to $e^+e^- \to e^-\barnu _e u \bard$ through
a $\gamma(Z)$ and $W$ bremsstrahlung followed by $\gamma(Z)+W \to W^*
\to u\bard$. In WTO the width of a vector boson is always zero for a
$t$-channel exchange thus for CC20 gauge invariance will be violated
because of the $W$ width in the $s$-channel. The {\it fermion-loop}
scheme includes in the calculation the {\it imaginary} part of the two
triangle diagrams (with opposite charge flow) with a fermionic loop
which represents the first order correction to the $VWW$ vertex. 
In the fermionic loop we include leptons, neutrinos and $u,d,c,s$ quarks,
i.e. everything which is allowed in the $W$ decay. In principle and for
$m_+^2 \geq (m_t+m_b)^2$ there are two additional imaginary parts for the 
vertex which correspond to cutting a $t$-line and a $b$-line. These
extra, $m_t$-dependent, contributions have not been included so far in
our calculation.
To be more specific we take into account the corrections to

\begin{eqnarray}
V_{\nu}(q)+W_{\mu}(p) &\to& W_{\alpha}(k), \qquad V = \gamma,Z  \nll
p &=& p_--q_1, \quad p = p_+-q_2, \quad k = q_3+q_4.
\end{eqnarray}

\noindent
The correction factor becomes

\begin{equation}
V^{\gamma(Z)}_{\mu\nu\alpha} = \lb 2\,\pi\rb^4\,i\,gs_{_W}\lb c_{_W}\rb\left[
-\frac{9}{16}\,i\,{{g^2}\over {\pi}}{\cal V}_{\mu\nu\alpha}\right],
\end{equation}

\noindent
where we have explicitly {\it factorized} the lowest order coupling. 
The corrected vertex $V$ does not factorize into the corresponding lowest 
order amplitude but instead three additional form factors must be included. 
They give

\begin{eqnarray}
{\cal V}_{\mu\nu\alpha} &=&
\sum_{i=1,4}\,C_iV^i_{\mu\nu\alpha},  \nll
C_i &=& {y\over {\lambda}}\left[\left(a_i + {{b_i}\over
{\lambda}} + \right)\,I_0 + \left(c_i + {{d_i}\over {\lambda}}\right)\,L_0
\right], \quad i= 1,\dots 3 \nll
C_4 &=& {y\over {\lambda}}\left[\left(a_4 + {{b_4}\over
{\lambda}} + {{e_4}\over{\lambda^2}}\right)\,I_0 + \left(c_4 + {{d_4}\over 
{\lambda}} + {{f_4}\over {\lambda^2}}\right)\,L_0\right].
\end{eqnarray}

\noindent
with

\begin{eqnarray}
V^1_{\nu\alpha\mu} &=& - \left[\delta_{\alpha\mu}k_{\nu}+
\delta_{\nu\mu}q_{\alpha}-\delta_{\nu\alpha}k_{\mu}\right],  \nll
V^2_{\nu\alpha\mu} &=& \left(\delta_{\nu\mu}q_{\alpha}-
\delta_{\nu\alpha}k_{\mu}\right),  \nll
V^3_{\nu\alpha\mu} &=& \left(\delta_{\nu\mu}q_{\alpha}+
\delta_{\nu\alpha}k_{\mu}\right),  \nll
V^4_{\nu\alpha\mu} &=& - {1\over {k^2}}\,k_{\nu}q_{\alpha}k_{\mu}.
\end{eqnarray}

\noindent
Moreover

\begin{eqnarray}
\lambda &=& \lb p\cdot q\rb^2 - p^2q^2,  \nll
I_0 &=& \frac{1}{2}, \qquad L_0 = {1\over {4\,\sqrt{\lambda}}}\,\log
{{p\cdot q - \sqrt{\lambda}}\over {p\cdot q + \sqrt{\lambda}}}.
\end{eqnarray}

\noindent
and

\begin{eqnarray}
a_1 &=& 6x-\frac{4}{3}z,   \nll
a_2 &=& 5x, \nll  
a_3 &=& -3x-\frac{2}{3}(y-z),   \nll
a_4 &=& \frac{8}{3}y, \nll
b_1 &=& -x(yz-4xy+xz),   \nll
b_2 &=& \frac{3}{2}x(-yz+4xy-xz), \nll
b_3 &=& \frac{1}{2}x(5yz-4y^2-8xy+xz),   \nll
b_4 &=& -\frac{1}{2}y(-\frac{176}{3}xy+16xz-16x^2+\frac{4}{3}yz),  \nll
c_1 &=& -4x(y-z+2x),  \nll
c_2 &=& 2x(-y+z-4x),  \nll
c_3 &=& 2x(3y-z+2x),  \nll
c_4 &=& -16xy,  \nll
d_1 &=& -x^2(-3yz+2y^2+6xy-xz),   \nll
d_2 &=& \frac{3}{2}x^2(3yz-2y^2-6xy+xz),  \nll
d_3 &=& \frac{1}{2}x(-2y^2z-9xyz+14xy^2+10x^2y-x^2z),  \nll
d_4 &=& -4xy(-3yz+2y^2+18xy-3xz+2x^2),   \nll
e_4 &=&  -5xy^2(yz+3xz-6xy-2x^2),   \nll
f_4 &=& -10x^2y^2(-2yz+y^2-2xz+6xy+x^2),  
\end{eqnarray}

\noindent
with

\begin{equation}
x = q^2, \qquad y = k^2, \qquad z= k^2+q^2-p^2.
\end{equation}

\noindent
We have verified that the inclusion of the full correction in WTO
accounts for an increase of the CPU time corresponding to a factor of
$1.3$, which we consider still very reasonable. It should be stressed
here that we include the {\it imaginary} part for both $\gamma WW$
and $ZWW$ vertices in order to preserve the full $SU(2)\otimes U(1)$
gauge invariance. This can be most easily seen by considering the
sub-process $e^-W^+ \to e^- u\bard$ and by writing the corresponding
Ward identity $p_{\mu}A^{\mu}_{_W}+i\,\xi^2\wm A_{\phi} = 0$ in the
$R_{\xi}$ gauge ($\phi$ being the Higgs ghost).
The relevance of the fermion-loop scheme for CC processes with electrons
in the final state becomes more and more pronounced for smaller and smaller
electron scattering angles.
We have explicitly verified that a selection cut of $10^o$ on the $e^-$ 
scattering angle in the process $e^+e^- \to e^-{\bar\nu}_e\nu_{\mu}\mu^+$
is enough to avoid any problem. Partial results are shown in table $2$,
where $\theta_m\angle (e^-,e^-)$ and we have defined $\delta_{_{FL}} = 1 -
\sigma/\sigma_{_{FL}}$.

\begin{table}[hbtp]
\begin{center}
\begin{tabular}{|c|c|c|c|}
\hline
$\theta_m\,$(deg) & $\sigma\,$(nb) & $\sigma_{_{FL}}\,$(nb) & 
$\delta_{_{FL}}(\%)$\\
\hline
$10^o$  & $0.20140(4)$  & $0.20143(4)$ &  $+0.01$ \\ 
$5^o$   & $0.20653(3)$  & $0.20643(3)$ &  $-0.05$ \\ 
$1^o$   & $0.21554(21)$ & $0.21014(3)$ &  $-2.57$ \\ 
\hline
\end{tabular}
\caption{\it Inclusion of the gauge restoring terms at $190\,$GeV in
$e^+e^- \to e^-{\bar\nu}_e\nu_{\mu}\mu^+$, for $\theta_m \leq \theta_{e^-} \leq
\pi-\theta_m$. In both cases the $W$ width is running. Moreover $E_{e,\mu} \geq
1\,$GeV and $\theta(e^-,\mu^+) \geq 5^o$.} 
\end{center}
\label{tab2}
\end{table}

\noindent
Needless to say the fully extrapolated cross section for $e^+e^- \to
e^-{\bar\nu}_e\nu_{\mu}\mu^+(u\bard)$ requires both gauge reparation
and non zero electron mass. Thus strictly speaking one cannot set
$\theta_m$ to zero in the present version of the program.
For a better understanding of the effect of restoring gauge invariance in CC20
we have shown in Fig. $1$ the angular distribution $d\sigma/d\cos\theta_l$ 
where $\theta_l$ is defined by

\begin{equation}
\ti = x_+e_1{{1+\cos\theta_l}\over {x_-+x_+-(x_--x_+)\cos\theta_l}},
\end{equation}

\noindent
i.e. $\theta_l$ is the angle between the incoming $e^-$ and the outgoing $l^-$.
In Fig. $1$ we have shown the angular distribution for 
$e^+e^- \to e^-\barnu_e\nu_{\mu}\mu^+$ with and without the fermion loop
corrections as compared with the similar distribution for a CC11 process
$e^+e^- \to \mu^-\barnu_{\mu}\nu_{\tau}\tau^+$.

%\begin{figure}[ht]
%\vspace{0.1cm}
%\centerline{
%\epsfig{figure=cc20.ps,height=12cm,angle=0}
%}
%\caption{\it The angular distribution for some CC20 process with and without
%FL-scheme.}
%\label{fig1}
%\end{figure}

\subsection{Final state QED radiation}

In this rather short section we would like to give an idea of the problematic
connected with QED radiation. Because of gauge invariance there is no
meaningful splitting of the QED corrections between initial state and
final state, splitting which was instead applicable for LEP~1 physics.
There are various attempts to circumvent the problem, most noticeably
the current-splitting of GENTLE~\cite{gen}, but as a matter of principle
even this technique cannot guarantee an unambiguous answer.

From this point of view only the full ${\cal O}(\alpha)$ calculation
makes sense and correctly reproduces the ${\cal O}(\alpha\times\,$const$)$
terms of the QED corrections. As in any other calculation of this type
we would end up with a correction $\delta_{QED}$ factor of the following
structure

\begin{equation}
\delta_{QED} = \delta_{soft} + \delta_{virt} + \delta_{hard},
\end{equation}

\noindent
after which the proper exponentiation can be performed. What we know of
$\delta_{QED}$ are the {\it universal}, leading logarithmic terms but
a calculation of the virtual corrections, including up to pentagon and
hexagon diagrams, is missing. In such a situation it is extremely risky
to partially account for final state QED radiation, even though any
{\it realistic} estimate of cross sections and of energy losses should
include at least some reasonable guess.

To this end we have preliminary investigated the inclusion of QED final
state radiation to CC11 processes. We have made use of our knowledge of
the universal {\it soft} terms and moreover of the general and simple
result of ref.~\cite{cgr} for the emission of hard collinear radiation by 
charged particles. The latter allows to include hard photons emitted within
a cone of fixed half-opening angle $\delta_c$, in the limit $\delta_c << 1$.
This result is based on a rigorous gauge invariant procedure at
$\delta_c << 1$ so that our correction factor for each charged fermionic
line will be

\begin{equation}
\delta_{_{FSR}} = \exp\lb {{\alpha}\over {\pi}}\delta_{soft}\rb\, \lb 1 +
{{\alpha}\over {\pi}}\,\delta_{coll}\rb.
\end{equation}

\noindent
In this way we can only make a very rough estimate of the effect of final state
radiation since all the {\it hard} constant from virtual corrections are
still missing. Moreover the procedure of exponentiation is also far from
unique since we could decide to exponentiate not only the leading logarithms
but also numerically relevant terms from $\delta_{coll}$.
Let $E_i,m_i$ be the energy and the mass of the emitting fermion, moreover we 
denote by $E_{_0} = e_{_0}\sqrt{s}$ the energy threshold, i.e $E \geq E_{_0}$.
In our formulation

\begin{equation}
E_i = \frac{1}{2}\eta_i\sqrt{s}, \quad \eta_i = x_+e_i+\lb x_+-x_-\rb t_i.
\end{equation}

\noindent
According to ref.~\cite{cgr} and to the subsequent generalization~\cite{cmn}
we define

\begin{eqnarray}
\rho_i &=& {{\delta_c\eta_i}\over {2\,m_i}},  \nll
1-\veps_i &=& {{2\,e_{_0}}\over {\eta_i}},
\end{eqnarray}

\noindent
and compute $\delta_{coll}$ as $Q_i^2\,C$ with $C$ given by eqs.(5-6) of
ref.~\cite{cmn}.

Our preliminary results show very little effect on the cross section for
$e^+e^- \to \mu^-{\bar\nu}_{\mu}u\bard$, for instance we obtain
$\sigma = 0.59192(4)\,$nb and $\sigma_{_{FSR}} = 0.59165(4)\,$nb at 
$\sqrt{s} = 190\,$GeV and for $\delta_c = 5^o$. The effect of QED final
state radiation must certainly be included for any reliable determination
of the physical observables at LEP~2 but we have decided for not including
it in the present version of WTO since a more detailed theoretical 
investigation is needed. 

\subsection{Program structure}

We start this section with a brief description of the general features
of the program.
With WTO it is possible to access $29$ out of $32$ 4-fermion (massless)
processes. The most relevant INPUT parameters are given in the BLOCK DATA INIT.
They are:

\begin{enumerate}

\item {\tt WM}. The $W$ mass (GeV).

\item {\tt ZM}. The $Z$ mass (GeV).

\item {\tt ZG}. The $Z$ total width (GeV).

\end{enumerate}

Other quantities like the $W$ width are not external input parameters but
rather they are computed internally as given by the standard model.
There is no possibility of changing the weak mixing angle independently
from the others input parameters, thus $s_{_W}^2$ is always a derived
quantity.
In addition to the $29$ processes which include gluon exchange in the
non-leptonic case there are $4$ processes ($e^+e^- \to \barf f \barb  b$)
which can be computed and where the Higgs boson exchange is included.
All processes receive and internal number according to the scheme given
in appendix.
Moreover each process belong to a particular class, but it is important
to know only the highest class in a particular family, thus
CC11, CC20, NC21, NC24, NC32, Mix43, NC48, NC64. 
WTO will always perform a call to the NAG routine X02AJF which returns
the machine precision ({\tt ZRM}). Some of the internal controls are based on 
this quantity so one should be aware of the fact that, for instance, 
on a VAXstation$\,\cdot\,$4000$\,\cdot\,$90 {\tt ZRM} = $0.14E-16$
while on a Alpha AXP-2100 {\tt ZRM} = $0.11E-15$.
It is perhaps appropriate to give here a brief summary of the computational
speed versus requested precision for some of the processes. In table $3$
we have indicated both entries for a process representative of each
family.

\begin{table}[hbtp]
\begin{center}
\begin{tabular}{|c|c|c|c|}
\hline
Family & Process & Rel. error & CPU time \\
\hline
CC11 & All & $\leq 0.01\%$ & $3:38:53$  \\
CC20 & All & $\leq 0.02\%$ & $2:42:57$  \\
NC24 & $\mu^+\mu^-\tau^+\tau^-$ & $0.25\%$ & $11:47:47$  \\
NC32 & ${\bar u}u{\bar c}c$ & $0.08\%$ & $12:30:52$  \\
NC19 & $\mu^+\mu^-\barnu _e\nu_e$ & $0.48\%$ & $12:06:14$  \\
Mix43 & $\mu^+\mu^-\barnu _{\mu}\nu_{\mu}$ & $0.08\%$ & $7:59:26$  \\ 
NC48 & $e^+e^- {\bar u}u$ & $0.17\%$ & $2:10:21$ \\
NC64 & $\barnu_{\mu}\nu_{\mu}\barnu_{\mu}\nu_{\mu}$ & $0.03\%$ & 
$5:24:42$  \\
\hline
\end{tabular}
\caption{\it CPU time needed on a Alpha AXP-2100 computer for a given relative
error (in percent) and for the some representative processes in each family.}
\end{center}
\label{tab3}
\end{table}

\noindent
As it can be seen from table $3$ there are considerable differences
in CPU time between different processes. The reason is not only connected with
the number of diagrams contributing but also with the number of helicity
sets required by each process. Always working in the massless fermion limit 
we have only $2(3)$ sets of non zero helicity states for CC11(CC20) but up
to $8(12)$ for NC24(NC48).

\subsection{Input}

The meaning of the input parameters is the following.
\vskip 15pt

\noindent
{\tt OPRT(CHARACTER*1)}

\noindent
There is the possibility of printing some additional information about
the calls to the various subroutines. This however requires a detailed
knowledge of the internal structure of the calculation. Thus, by default,
{\tt OPRT}='N' and it can be set to 'Y' only by changing the corresponding
line in BLOCK DATA {\tt INIT}.
\vskip 15pt

\noindent
{\tt NPTS(INTEGER*4)}

\noindent
The actual number of points for the integration. Built-in choices are {\tt 
NPTS}=$1-10$. If more integration points are needed then the array {\tt 
VK(NDIM)} must be re-initialized through a call to the NAG routine D01GZF.
The optimal coefficients for $p$-point integration over the $n$-cube $[0,1]^n$
require that $p$ is a prime number or $p$ is a product of $2$ primes, $p_2$
and $p_1$ chosen  to be the nearest prime integer to $p_2^2$.
The built-in choices are shown in table $4$.

\begin{table}[hbtp]
\begin{center}
\begin{tabular}{|c|c|}
\hline
{\tt NPTS} & $p = p_1\times p_2$ \\
\hline
$1$ & $2129$ \\
$2$ & $5003$ \\
$3$ & $10007$ \\
$4$ & $20011$ \\
$5$ & $40009$ \\
$6$ & $80021$ \\
$7$ & $99991$ \\
$8$ & $10193\times 101$ \\
$9$ & $22807\times 151$ \\
$10$ & $32771\times 181$ \\
\hline
\end{tabular}
\caption{\it Built-in choices for $p$, needed for $p$-point integration over 
the $n$-cube $[0,1]^n$.}
\end{center}
\label{tab4}
\end{table}

\vskip 15pt
\noindent
{\tt NRAND(INTEGER*4)}

\noindent
The number of times where the integral is computed in order to give an 
estimate of the standard error (usually {\tt NRAND} $\leq 6$).
A `fast' estimate of the result with standard error around $1\%$ is
already achieved with {\tt NPTS}=$4$, {\tt NRAND}=$4$. A more accurate but still
intermediate answer will require ($8,6$) and a very precise, but also
very time consuming, estimate will proceed with ($10,6$). There is no
intrinsic limit to {\tt NRAND} while {\tt NPTS} $\leq 10<$, unless an
independent call to NAG routine D01GZF is performed.

\vskip 15pt
\noindent
{\tt IPR(INTEGER*4)}

\noindent
The catalog number of the process [$1-33$]. The internal structure 
of WTO is organized in such a way that the user does not have
to bother about quantum numbers of the final state. 

\vskip 15pt
\noindent
{\tt IPR0(INTEGER*4)}

\noindent
If {\tt IPR} $\leq 3$ there is the additional possibility of taking into
account only the CC03 part of the CC11 family.

\vskip 15pt
\noindent
{\tt RS(REAL*8)}

\noindent
The c.m.s energy in GeV.

\vskip 15pt
\noindent
{\tt OPEAK(CHARACTER*1)}

\noindent
In general we are dealing with double resonant, single resonant and 
non-resonant diagrams. {\tt OPEAK}='Y[N]' will select the corresponding 
mapping. 

Let us assume that we want to reach the same level of accuracy for
all possible cases. As a consequence we observe that different processes 
under different sets of kinematical cuts will require values of
{\tt NPTS} and {\tt NRAND} which may differ substantially. To illustrate this
point we consider a particular example, $e^+e^- \to \mu^+\mu^-\tau^+\tau^-$.
Moreover we only include simple kinematical cuts on the invariant mass of the
two fermion-antifermion pair, which can be coupled to a photon. The 
{\it background} to $Z-Z$ production is therefore large, expecially when
we allow for small cuts, $M(\mu^+\mu^-),M(\tau^+\tau^-) \geq M_{_0}$ with
$M_{_0}$ much smaller than $\zm$. To show our point we fix {\tt NPTS}=$8$
and {\tt NRAND}=$6$ and vary $M_{_0}$. Table $5$ clearly
illustrates as the numerical accuracy becomes better and better for
growing $M_{_0}$ and fixed number of points.

\begin{table}[hbtp]
\begin{center}
\begin{tabular}{|c|c|c|}
\hline
$M_{_0}\,$(GeV) & $\sigma\,$(fb) & Rel. error($\%$) \\
\hline
$5$  & $10.29(8)$ & $0.77$ \\
$10$ & $6.61(3)$  & $0.43$ \\ 
$30$ & $3.130(2)$ & $0.07$ \\
$50$ & $2.226(1)$ & $0.03$ \\
\hline
\end{tabular}
\caption{\it The effect of varying the cut on fermion-antifermion pair in
$e^+e^- \to \mu^+\mu^-\tau^+\tau^-$, inclusive of QED radiation.} 
\end{center}
\label{tab5}
\end{table}

\noindent
In this example we have used {\tt OPEAK}='N' for $M_{_0} = 5,10\,$GeV
and {\tt OPEAK}= 'Y' for $M_{_0} = 30,50\,$GeV. High precision for very
loose cuts requires therefore a much higher number of points while it
is considerably easier to reach $0.1\%$ or better for more {\it realistic}
cuts. It goes without saying that a more stringent cut on the
uninteresting boundaries of the phase space implies high precision achievable 
with high computational speed. Finally we observe that for some of the
processes with very loose cuts a repetition with an increasing sequence
of high values of {\tt NPTS} can eventually yield erratic results.

\vskip 15pt
\noindent
{\tt ALS(REAL*8)}

\noindent
For NC32 and non-leptonic NC64, Mix43 the value of $\alpha_s(\wm)$ must be 
initialized. Moreover for CC processes it is possible to include NQCD effects
(Naive QCD). This is controlled by the variable OQCD.

\vskip 15pt
\noindent
{\tt OQCD(CHARACTER*1)}

\noindent
Inclusion of NQCD both in the total $W$ width and in the final state, {\tt ALS} 
must be initialized. Thus

\begin{equation}
\Gamma_{_W} = \Gamma_{_W}^0\,\lb 1 + \frac{2}{3}{{\alpha_s}\over {\pi}}\rb.
\end{equation}

\vskip 15pt
\noindent
{\tt OFL(CHARACTER*1)}

\noindent
Whenever a CC20 process is considered an option is available for introducing
the gauge restoring terms according to the FL-scheme. The $W$-width is 
always taken to be {\it running}. The $e^-$ scattering angle cannot be set
to zero even in the presence of gauge reparation due to the approximation
of $m_e = 0$.

\vskip 15pt
\noindent
{\tt OFSR(CHARACTER*1)}

\noindent
In the present version of the program this flag has been set to its
default, {\tt OFSR}='N' and it should not be changed. It is responsible
for the inclusion of QED final state radiation for CC11 processes, a branch
which deserves a more detailed theoretical analysis.  

\vskip 15pt
\noindent
{\tt ITCM(INTEGER*4)}

\noindent
For CC11 and CC20 several observables can be computed:

\begin{itemize}

\item 0 The cross-section 
\item 1 unrenormalized moments of 

\begin{equation}
E_{\gamma} = \lb 1 - {{x_++x_-}\over 2}\rb\,\sqrt{s}.
\end{equation}

\item 2 for CC10. Unrenormalized moments of $E_{\mu}$ as defined in
eq.(\ref{ps}).

\item 3 For CC10. Unrenormalized $T_n(\cos\theta_{\mu})$, where
$T_n$ is the $n$-th Chebyshev polynomial and $\cos\theta_{\mu}$ is defined in
eq.(\ref{ps}) with $i=1$.
 
\item 4 unrenormalized $T_n(\cos\theta_{W^-})$. The $W^-$ scattering
angle is defined by

\begin{eqnarray}
\cos\theta_{W^-} &=& {1\over {\beta_{_{W^-}}}}\,\lb 1 - x_-{{\tw}\over 
{E_{_{W^-}}}} \rb,  \nll
E_{_{W^-}} &=& \frac{1}{2}\,\left[x_+\lb 1 + m_-^2 - m_+^2\rb -
\lb x_+ - x_-\rb\tw\right],  \nll
\beta_{_{W^-}}^2 &=& 1 - {{v m_-^2}\over {E_{_{W^-}}^2}}.
\end{eqnarray}

\item 5 unrenormalized $T_n(\cos\theta_{W^+})$. The $W^+$ scattering
angle is defined by

\begin{eqnarray}
\cos\theta_{W^+} &=& {1\over {\beta_{_{W^+}}}}\,\lb 1 - x_-{{1-\tw}\over 
{E_{_{W^+}}}} 
\rb,  \nll
E_{_{W^+}} &=& \frac{1}{2}\,\left[x_+\lb 1 + m_+^2 - m_-^2\rb -
\lb x_+ - x_-\rb(1-\tw)\right],  \nll
\beta_{_{W^+}}^2 &=& 1 - {{v m_+^2}\over {E_{_{W^+}}^2}}.
\end{eqnarray}

\item 6 unrenormalized $W$ virtuality defines as 

\begin{equation}
V_{_W} = \sqrt{v}\lb m_-+m_+\rb - 2\,{{\wm}\over s}.
\end{equation}

\item 7 $M(W^+)+M(W^-)$ distribution 
\item 8 $M(W^-)-M(W^+)$ distribution 
\item 9 $M(W^+)$ distribution 

\item 10 $c = \cos\theta(l^-)$ distribution where

\begin{equation}
\ti = x_+e_1{{1+c}\over {x_-+x_+-(x_--x_+)c}},
\end{equation}

\noindent
and $theta$ is the angle between the incoming $e^-$ and the outgoing $l^-$.

\end{itemize}

\noindent
while for all other classes only the cross-section will be available.
Please consult the WW/eg report~\cite{wweg} for more detailed informations 
concerning this quantities. Here unrenormalized means not divided by the 
corresponding cross-section.

\vskip 15pt
\noindent
{\tt ITCNM(INTEGER*4)}

\noindent
If $1 \leq $ {\tt ITCM} $\leq 6$ the order of the moments can be chosen.

\vskip 15pt
\noindent
{\tt DIST(REAL*8)}

\noindent
For {\tt ITCM}=$7,8,9$ gives the value of $M_{W^+}+M_{W^-}$, $M_{W^-}-M_{W^+}$ 
or $M_{W^+}$ at which the differential cross section is required.
For {\tt ITCM}=$10$ {\tt DIST} is $\theta \angle (e^-,l^-)$.
\vskip 15pt

\noindent
Moreover for {\tt ITCM=}$0$ there is the additional possibility of
{\it binning} the $M(W^-)\pm M(W^+)$ distributions by requiring the
calculation of the cross-section with $M_{_1} \leq  M(W^-)\pm M(W^+) \leq 
M_{_2}$.

\vskip 15pt
\noindent
{\tt OBIN(CHARACTER*1)}

\noindent
{\tt OBIN}='N'['P','M'] selects no binning or $\pm$ binning. For the latter
we initialize

\vskip 15pt
\noindent
{\tt ABP(M),BBP(M)(REAL*8)}

\noindent
the limits $M_{_1},M_{_2}$.

\vskip 15pt
\noindent
{\tt OCOUL(CHARACTER*1)}

\noindent
For CC processes the Coulomb correction factor can be included.

\vskip 15pt
\noindent
{\tt IOS(INTEGER*4)}

\noindent
Refers to the choice of the weak mixing angle, $s_{_W}^2$ and of the $SU(2)$
coupling constant $g$. The user is advised to adopt {\tt IOS=}$1$ which is 
the choice of a large variety of programs and corresponds to ($\gf$ being
the Fermi coupling constant)

\begin{eqnarray}
s_{_W}^2 &=& {{\pi\alpha}\over {{\sqrt 2}\gf\wm^2}},  \nll 
g^2 &=& {{4\pi\alpha}\over {s_{_W}^2}}
\end{eqnarray}

\noindent
although a more sensible choice (connected with the Ward identities) is
given by

\begin{eqnarray}
s_{_W}^2 &=& 1 - {{\wm^2}\over {\zm^2}}, \nll
g^2 &=& 4{\sqrt 2}\gf\wm^2. 
\end{eqnarray}

\vskip 15pt
\noindent
{\tt ORAL(CHARACTER*1)}

\noindent
If we adopt {\tt IOS=}$1$ then the scale must be chosen for evaluating
$\alpha$. If {\tt ORAL}='F' then it is possible to enter any value
for {\tt ALWI}= $1/\alpha$ while for {\tt ORAL}='R' the program will
compute $\alpha(s)$ with a call to subroutine {\tt PSELF} and {\tt HADR5}.
The two routines evaluate {\tt DER}= {\tt DERL}+{\tt DERH}, {\tt DERL}
being the {\it perturbative} contribution to $\Delta r$ from leptons
and from the top quark while {\tt DERH} evaluates the light
hadron contribution using fits to the QED vacuum polarization from
$e^+e^-$ data (courtesy of F.~Jegerlehner~\cite{jeg}).

\begin{equation}
\alpha(s) = {{\alpha(0)}\over {1-\Delta r}}.
\end{equation}

\vskip 15pt
\noindent
{\tt IOSF(INTEGER*4)}

\noindent
Refers to the choice of the Structure Function for Initial State
QED radiation. Born observables require {\tt IOSF}=$0$, for the
corrected ones the user is advised to have {\tt IOSF}=$1$. Indeed this choice
is the {\it default} adopted by the WW/eg working group
and corresponds to having $D(x,s)$ of eq.(~\ref{sf}) entirely in terms
of $\beta$~\cite{sf1},

\begin{equation}
D_r(x,s)= - \frac{1}{2}\lb 1-x\rb^{-\beta/2}\,\lb 1-x^2\rb
+ {\cal O}\lb \beta\rb.
\end{equation}

\noindent
Instead {\tt IOSF}=$2$ corresponds to the so-called $\eta$-scheme~\cite{sf2}

\begin{eqnarray}
D_r(x,s) &=& - \frac{1}{2}{\eta\over \beta}\lb 1-x\rb^{-\beta/2}\,\lb 1-x^2\rb
+ {\cal O}\lb {{\eta^2}\over {\beta}}\rb,  \nll
\eta &=& 2\,{\alpha\over \pi}\,\log{s\over {m_e^2}},
\end{eqnarray}

\noindent
which respects gauge invariance. Finally {\tt IOSF}=$3$ gives a mixed 
treatment as described in~\cite{sf3}.
There is no way in which one can present any {\it realistic} estimate
of the theoretical error for $4$-fermion production with the actual level
of knowledge that we have. The only {\it pragmatic} alternative is to
examine the variation induced in our results by changing RS and SF.
The first option roughly simulates (probably {\it underestimate}) the
uncertainty connected with our ignorance of the ${\cal O}(\alpha)$ 
electroweak corrections while the second one reflects the uncertainty
related to our ignorance about the complete ${\cal O}(\alpha\times\,$const) QED
corrections which, at the moment, only control correctly the leading
logarithmic parts. This we have done by allowing {\tt IOS}=$1,2$ and
{\tt IOSF}=$1,2,3$ in the calculation of the {\it representative} process
$e^+e^- \to \mu^-\barnu_{\mu},\nu_{\tau}\tau^+$. Results for the cross section
at $\sqrt{s} = 190\,$GeV and for canonical cuts are show in table $6$.

\begin{table}[hbtp]
\begin{center}
\begin{tabular}{|c|c|c|c|}
\hline
{\tt IOS}/{\tt IOSF} & 1 & 2 & 3  \\
\hline
1 & $0.19411(2)$ & $0.19395(2)$ & $0.19464(2)$ \\
2 & $0.19394(2)$ & $0.19378(2)$ & $0.19447(2)$  \\
\hline
\end{tabular}
\end{center}
\caption{\it $\sigma\,$(nb) for $e^+e^- \to \mu^-{\bar\nu}_{\mu},\nu_{\tau}
\tau^+$ at $190\,$GeV with $E_{\mu,\tau} \geq 1\,$GeV and
$10^o \leq \theta_{\mu,\tau} \leq 170^o$. Moreover $\theta_{\mu,\tau} 
\geq 5^o$.}
\label{tab6}
\end{table}

\noindent
Thus {\it naively} one could fix the central value at {\tt IOS}=$2$ and 
{\tt IOSF}=$2$ which preserve gauge invariance and assign a quite
asymmetric {\it theoretical error} of

\begin{equation}
\sigma = 0.19378 \pm 0.00002({\hbox{num.}})^{+0.00086}_{-0.00000}
({\hbox{theor.}}).
\end{equation}

\noindent
Moreover we have also show in tables $7-8$ the effect of 
varying {\tt IOS/IOSF} in the process $e^+ e^- \to \mu^-{\bar\nu}_{\mu}u
{\bar d}$ both for $\sigma$ and for $<E_{\gamma}>$. In this case we have 
included the Coulomb correction factor, {\tt OCOUL}='Y', and the NQCD 
corrections, {\tt OQCD}='Y'. Canonical cuts are applied.
The largest uncertainty for $<E_{\gamma}>$ is of $1.9$, $3.2$, $9.9$, $20.5\,$
MeV for $E_{cm} = 161, 175, 190, 205\,$GeV respectively.

\begin{table}[hbtp]
\begin{center}
\begin{tabular}{|c|c|c|c|c|c|c|}
\hline
$E_{cm}\,$(GeV)/IOS-IOSF & 1-1 & 1-2 & 1-3 & 2-1 & 2-2 & 2-3 \\
\hline
161 & 0.4688 & 0.4673  & 0.4674  & 0.4685  & 0.4669  & 0.4670 \\
175 & 1.1250 & 1.1219  & 1.1221  & 1.1251  & 1.1220  & 1.1222  \\
190 & 2.1579 & 2.1484  & 2.1489  & 2.1583  & 2.1488  & 2.1493  \\
205 & 3.2317 & 3.2119  & 3.2129  & 3.2324  & 3.2126  & 3.2135  \\
\hline
\end{tabular}
\end{center}
\caption{\it $<E_{\gamma}>\,$(GeV) for $e^+ e^- \to \mu^-{\bar\nu}_{\mu}u
{\bar d}$ with $E_{\mu} \geq 1\,$GeV, $E_{u,d} \geq 3\,$GeV, $M_{ud} \geq
5\,$GeV, $10^o \leq \theta_{\mu} \leq 170^o$, $\theta_{\mu,u},\theta_{\mu,d}
\geq 5^o$. }
\label{tab7}
\end{table}

\begin{table}[hbtp]
\begin{center}
\begin{tabular}{|c|c|c|c|c|c|c|}
\hline
$E_{cm}\,$(GeV)/IOS-IOSF & 1-1 & 1-2 & 1-3 & 2-1 & 2-2 & 2-3 \\
\hline
161 &0.13206 &0.13204 &0.13250 &0.13201 &0.13198 &0.13244 \\
175 &0.49207 &0.49186 &0.49358 &0.49177 &0.49156 &0.49329 \\
190 &0.60240 &0.60192 &0.60404 &0.60188 &0.60139 &0.60352 \\
205 &0.62828 &0.62754 &0.62977 &0.62764 &0.62691 &0.62913 \\
\hline
\end{tabular}
\end{center}
\caption{\it $\sigma\,$(nb) for $e^+ e^- \to \mu^-{\bar\nu}_{\mu}u
{\bar d}$. The cuts are the same as in table~ref{tab7}.}
\label{tab8}
\end{table}

\vskip 15pt
\noindent
{\tt OCUTS(CHARACTER*2)} 
It is the main decision branch for kinematical cuts. 

\noindent
There are few internal options.

\begin{enumerate}

\item {\tt OCUTS}='EX'. Fully extrapolated set-up, use it only for processes
                  free of singularities, e.g. never for CC20.
\item {\tt OCUTS}='CC'. The Canonical Cuts used within the WW/eg Working Group.
                  In this case cuts are automatically selected. Canonical
                  Cuts are also available for NC21,NC25 and NC50 processes.
                  For completeness we make a list of the chosen canonical
                  cuts~\cite{wweg}. Let $l$ be any final state charged lepton
                  and $q$ any final state quark. Then

\begin{itemize}

\item $E_l \geq 1\,$GeV, $E_q \geq 3\,$GeV.

\item $M(q_i,q_j), M(q_i,{\bar q}_j), M({\bar q}_i,{\bar q}_j) \geq 5\,$GeV.

\item $10^o \leq \theta_l \leq 170^o$.

\item $\theta(l_i,q_j), \theta(l_i,{\bar q}_j) \geq 5^o$.

\end{itemize}

\item {\tt OCUTS}='SC'. No cuts at all but whenever a $\barf f$ pair appears
                  in the final state a cut is applied to prevent singularity 
                  due to $\gamma^*\to \barf f$.
                  Again not to be used for CC20 etc. Actually it can
                  ONLY be used for CC11, NC19(NC21), NC24(NC25), NC50 processes.
                  The `security' cut for CC11, NC19 and NC24 can be 
                  initialized through the variable {\tt STHG}. For Higgs 
                  physics, NC21, NC25 and NC50 processes, the `simple' cuts 
                  can be initialized through the variables {\tt ZCUT, BCUT}.

\item {\tt OCUTS}='FC'. You adopt your own cuts. Several variables must be 
                  initialized. The important thing to pay attention to is
                  the `ordering' of the final states. One should have in
                  mind the classification given in appendix, therefore if 
                  {\tt IPR}=18
                  is selected one must attribute cuts according to the
                  fact that $1=u, 2={\bar u}, 3=\nu_e, 4=\barnu _e$.
                  The internal coding will attribute the following labels:
                  for CC(Mix) processes $e^+e^- \to d(1) {\bar u}(2) u'(3) 
                  {\bar d'}(4)$, where $u= \nu_l,u,c$ and $d = l,d,s,b$. 
                  For NC processes $e^+e^- \to f(1) \barf (2) f'(3) 
                  \barf '(4)$. Boundaries for the scattering angles or 
                  final state energies are  stored as $1-4$.
                  Boundaries for final state angles $\theta_{ij}, i,i=1,4$ 
                  or final state invariant masses $M_{ij}, i,i=1,4$ are 
                  stored as a vector of dimension $6$, with entries 
                  $1-2/1-3/1-4/2-3/2-4/3-4$.

\end{enumerate}

\vskip 15pt
\noindent
{\tt STHG(REAL*8)}

\noindent
Selects $M(\barf f) \geq\,${\tt STHG}.

\vskip 15pt
\noindent
{\tt ZCUT,BCUT(REAL*8)}

\noindent
In $e^+e^- \to \barf f b \barb $ selects $\zm-\,${\tt ZCUT} $\leq M(\barf f) 
\leq \zm+\,${\tt ZCUT} and $M(\barb b) \geq\,${\tt BCUT}.

\vskip 15pt
\noindent
{\tt AIMM(I)(REAL*8)} 

\noindent
Min cuts in GeV on FS invariant masses $1-6$.

\vskip 15pt
\noindent
{\tt BIMM(I)(REAL*8)} 

\noindent
Max cuts in GeV on FS invariant masses $1-6$.

\vskip 15pt
\noindent
{\tt AEM(I)(REAL*8)}

\noindent
Lower cuts on FS energies (GeV) $1-4$.

\vskip 15pt
\noindent
{\tt ASAM(I)(REAL*8)} 

\noindent
Min cuts on the cosine of the scattering angles $1-4$.

\vskip 15pt
\noindent
{\tt BSAM(I)(REAL*8)} 

\noindent
Max cuts on the cosine of the scattering angles $1-4$.

\vskip 15pt
\noindent
{\tt AFSAM(I)(REAL*8)} 

\noindent
Min cuts on the cosine of the FS angles $1-6$.

\vskip 15pt
\noindent
{\tt BFSAM(I)(REAL*8)} 

\noindent
Max cuts on the cosine of the FS angles $1-6$.

\vskip 20pt
\noindent
Whenever the Higgs signal has to be investigated few additional flags
must be initialized. They are:

\vskip 15pt
\noindent
{\tt HMI(REAL*8)}

\noindent
The Higgs boson mass in GeV.

\vskip 15pt
\noindent
{\tt ALSM(REAL*8)}

\noindent
The value of $\alpha_s(\wm)$.

\vskip 15pt
\noindent
{\tt OQCD(CHARACTER*1)}

\noindent
The option for final state `naive' QCD corrections.

\vskip 15pt
\noindent
{\tt OGLU(CHARACTER*1)}

\noindent
The inclusion of the $H \to gg$ channel in the total Higgs width. 
The most {\it complete} treatment will therefore evolve $\alpha_s$ to the
scale $\mu = \hm$, evaluate the running $b,c$-quark masses and compute

\begin{eqnarray}
\Gamma_{_H} &=& {{G_G\hm}\over {4\,\pi}}\,\left\{ 3\,\left[ 
m_b^2(\hm) + m_c^2(\hm)\right]\,\left[ 1 + 5.67\,{\alpha_s\over \pi}
+ 42.74\,\lb{\alpha_s\over \pi}\rb^2\right] + m_{\tau}^2\right\} +
\Gamma_{gg},  \nll
\Gamma_{gg} &=& {{\gf\hm^3}\over {36\,\pi}}\,{{\alpha_s^2}\over {\pi^2}}\,
\lb 1 + 17.91667\,{\alpha_s\over\pi}\rb.
\end{eqnarray}

\vskip 15pt
\noindent
Again the `ordering' of final states is extremely important when one chooses
cuts. Please notice that for internal reasons WTO attributes the
following labels to {\tt IPR}=30-33: $1=b, 2=\barb , 3=\nu_l/l^-, 
4=\barnu_l/l^+$ but $1=\mu^-, 2=\mu^+, 3=b, 4=\barb $.

\subsection{Test run output}

The typical calculations that can be performed with the program are
illustrated in three different examples.

\vskip 15pt
\noindent
{\it Sample 1.} Evaluation of the total cross section for the
process $e^+e^- \to \mu^-, \barnu _{\mu}, u, {\bar d}$ inclusive
of QED initial state radiation.

\vskip 15pt
\noindent
{\it Sample 2.} Evaluation of the total cross section for the
process $e^+e^- \to e^+ e^- {\bar d} d$ without QED initial state radiation.

\vskip 15pt
\noindent
{\it Sample 3.} Evaluation of the total cross section for the
process $e^+e^- \to \mu^+\mu^-\barnu_{\mu}\nu_{\mu}$ with QED initial state 
radiation.

\vskip 15pt
\noindent
In sample $1-3$ we choose $E_{cm} = 190\,$GeV, $\wm = 80.23\,$GeV,
$\zm = 91.1888\,$GeV and $\Gamma_{_Z} = 2.4974\,$GeV. The various options are 
as follows. 

\noindent
For sample $1$ we have:

\vskip 15pt
\noindent
{\tt OPEAK='Y'}

\noindent
The double-resonating mapping is requested.

\vskip 15pt
\noindent
{\tt ITC=0}

\noindent
The total cross section is computed.

\vskip 15pt
\noindent
{\tt OCOUL='N'}

\noindent
The Coulombic correction factor is not included.

\vskip 15pt
\noindent
{\tt IOS = 1, IOSF = 1}

\noindent
The default choices for the renormalization scheme and the structure functions.

\vskip 15pt
\noindent
{\tt ORAL='F', ALWI = 128.07D0}

\noindent
We have fixed $\alpha$ to be $1/\alpha = 128.07D0$.

\vskip 15pt
\noindent
{\tt OCUT = 'CC'}

\noindent
Canonical Cuts are applied.

\vskip 15pt
\noindent
{\tt IPR = 2}

\noindent
The internal process number is selected.

\vskip 20pt
\noindent
For sample $2$ we have:

\vskip 15pt
\noindent
{\tt OPEAK='N'}

\noindent
The double-resonating mapping is NOT requested.

\vskip 15pt
\noindent
{\tt OCOUL='N'}

\noindent
The Coulombic correction factor does not apply for this process.

\vskip 15pt
\noindent
{\tt IOS = 1, IOSF = 0}

\noindent
The default choice for the renormalization scheme and NO QED radiation.

\vskip 15pt
\noindent
{\tt ORAL='F', ALWI = 128.07D0}

\noindent
We have fixed $\alpha$ to be $1/\alpha = 128.07D0$.

\vskip 15pt
\noindent
{\tt OCUT = 'CC'}

\noindent
Canonical Cuts are applied.

\vskip 15pt
\noindent
{\tt IPR = 25}

\noindent
The internal process number is selected.

\vskip 20pt
\noindent
For sample $3$ we have:

\vskip 15pt
\noindent
{\tt OPEAK='Y'}

\noindent
The double-resonating mapping is requested.

\vskip 15pt
\noindent
{\tt OCOUL='N'}

\noindent
The Coulombic correction factor is NOT not requested.

\vskip 15pt
\noindent
{\tt IOS = 1, IOSF = 1}

\noindent
The default choices for the renormalization scheme and the structure functions.

\vskip 15pt
\noindent
{\tt ORAL='F', ALWI = 128.07D0}

\noindent
We have fixed $\alpha$ to be $1/\alpha = 128.07D0$.

\vskip 15pt
\noindent
{\tt OCUT = 'CC'}

\noindent
Canonical Cuts are applied.

\vskip 15pt
\noindent
{\tt IPR = 20}

\noindent
The internal process number is selected.

\subsection{Program implementation}

In this section a short description of each routine implemented in WTO is given.

\begin{quote}{\footnotesize \begin{verbatim}

      BLOCK DATA INIT
      IMPLICIT REAL*8 (A-H,O-Z)
      CHARACTER*1,OPRT
*
      COMMON/PRT/OPRT
      COMMON/PARAM/EPS,QDELTA
      COMMON/FMASS/EM,RMM,TM,TQM,UQM,DQM,CQM,SQM,BQM
      COMMON/BPAR/WM,ZM,ZG,GF,PI,CFCT,FCNT,GE,ALPHAI,ALWI
*
*-----OFFICIAL SET OF DATA INPUT
*
      DATA WM/80.23D0/,CFCT/0.38937966D9/,FCNT/81.D0/,
     #     GF/8.2476172696D-6/,ZM/91.1888D0/,
     #     EM/0.51099906D-3/,PI/3.141592653589793238462643D0/,
     #     GE/0.5772156649D0/,ZG/2.4974D0/,TM/1.7771D0/,
     #     BQM/4.7D0/,CQM/1.55D0/,ALPHAI/137.0359895D0/
      DATA EPS/1.D-37/,QDELTA/0.D0/
      DATA RMM/0.10565839D0/,UQM/0.041D0/,DQM/0.041D0/,SQM/0.15D0/
      DATA TQM/175.D0/
      DATA OPRT/'Y'/
*
      END

\end{verbatim}}\end{quote}

\noindent
Masses and numerical constants are initialized. As it can be seen the top 
quark mass is also initialized. At the moment this is only due to the
following facts. Whenever $\alpha$ is evoluted at the scale $s$ then the
top quark enters the photon two point function. Morever in NC21, i.e.
$e^+e^- \to b\barb\nu_e\barnu_e$, there is a multiperipheral diagram whith
an internal top line.

\begin{quote}{\footnotesize \begin{verbatim}

      SUBROUTINE TOW(RS,IACM,AIMM,BIMM,AEM,ISAAM,ISABM,ASAM,BSAM,
     #               AFSAM,BFSAM,OTYPEM,CHUM,CHUPM,CHDM,CHDPM,CHFM,
     #               CHFPM,TIFM,TIFPM,FCUM,FCDM,ITCM,IPM,IRM,
     #               ITCNM,OXCM)
      IMPLICIT REAL*8 (A-H,O-Z)
*
      CHARACTER*1,OXCM,OQCD,OGLU,OFL,OCOUL,OFSR,OTRANS,OPRT
      CHARACTER*2,OFS
      CHARACTER*5 RLABS(3)
      CHARACTER*7 CLABS1(4),CLABS2(4),CLABS3(4),CLABS4(4),
     #            CLABS5(4),CLABS6(4),CLABS7(4),CLABS8(4),
     #            CLABS9(4),CLABS10(4),CLABS11(4),CLABS12(4),
     #            CLABS13(4),CLABS14(4),CLABS15(4),CLABS16(4),
     #            CLABS17(4),CLABS18(4),CLABS19(4),CLABS20(4),
     #            CLABS21(4),CLABS22(4),CLABS23(4),CLABS24(4),
     #            CLABS25(4),CLABS26(4),CLABS27(4),CLABS28(4),
     #            CLABS29(4),CLABS0(4)
      CHARACTER*4,OTYPEM,OTYPE
*
      PARAMETER (NOUT=99,NDIMMX=9)
*
      COMMON/MP/ZRM
      COMMON/FS/OFS
      COMMON/GG/OGLU
      COMMON/FLS/OFL
      COMMON/QCD/ALS
      COMMON/NPR/IPR
      COMMON/HIGGS/HM
      COMMON/PRT/OPRT
      COMMON/FSR/OFSR
      COMMON/OPA/DELC
      COMMON/DIS/DIST
      COMMON/KOUNT/IK
      COMMON/SP/PSG(4)
      COMMON/ISTRF/ISF
      COMMON/AQCD/OQCD
      COMMON/BME/BFACT
      COMMON/COUL/OCOUL
      COMMON/CHI/HCH(36)
      COMMON/TC/ITC,ITCN             
      COMMON/SHEL/OTRANS
      COMMON/IPT/IFZ(44)
      COMMON/ICUTS/IAC(4)
      COMMON/RM/RBM2,RCM2
      COMMON/TOPT/IOS,IOSF
      COMMON/ISA/ISAA,ISAB
      COMMON/OCHANNEL/OTYPE
      COMMON/NCC/CHF2,CHFP2,CONC(10)
      COMMON/CCLR/VUPL,VUPR,VDPL,VDPR
      COMMON/NCLR/VEL,VER,VELR,VFL,VFR,VFPL,VFPR
      COMMON/CCHANNEL/CHU,CHUP,CHD,CHDP,FCUC,FCDC
      COMMON/NCHANNEL/CHF,CHFP,TIF,TIFP,FCUN,FCDN
      COMMON/FMASS/EM,RMM,TM,TQM,UQM,DQM,CQM,SQM,BQM
      COMMON/HAPAR/RHM,RHM2,RHG,RHMG,SHG,SHGS,OPSHGS
      COMMON/EE/QCH,QCH2,VQR,VQL,HBE(24),HBO(24),HMP(24)
      COMMON/BPAR/WM,ZM,ZG,GF,PI,CFCT,FCNT,GE,ALPHAI,ALWI
      COMMON/CPAR/ALPHA,HBET,HBETI,OMHB,EOB,D0GL,G8,TFACT,PIH,ALW,
     #            ETA,FETA,BETA,G2
      COMMON/ACCHANNEL/OMCHU,OPCHU,OMCHUP,OPCHUP,OMCHDP,OPCHDP,
     #                 OMCHD,OPCHD,HCHUP,HCHU,HCHDP,HCHD
      COMMON/APAR/ARS,S,RWM,RWM2,RWG,RWMG,SWG,SWGS,OPSWGS,STH2,CTH2,
     #            HSTH2,TSTH2,SCTH2,ASTH2,TTH2,RZM,RZM2,RZG,RZMG,SZG,
     #            SZGS,OPSZGS,STH4,CTH4,VE,VF,VFP,RBQM2,RSZW,RSZW2,
     #            S0W,S0Z
      COMMON/SUBREG/DSM,USM,DSP,USP,RL(6),RR(6),SRL(6),SDSM,SDSP,VVL1,
     #              VVL2,VVL3,UL,OMUL,SUML
      COMMON/CUTS/AIM(6),BIM(6),AE(4),ASA(4),BSA(4),AFSA(6),BFSA(6),
     #            OMBSA(4),OPBSA(4),TEQ,RAE(4),OMASA(4),OPASA(4),
     #            SG12,CG12,SG13,CG13,SG14,CG14,SG23,CG23,SG24,
     #            CG24,SG34,CG34,SCT120,SCT130,SCT140,SCT230,
     #            SCT240,SCT340,SGAM(4),CGAM(4)

*
      DIMENSION VK(NDIMMX)
      DIMENSION IACM(4),AIMM(6),BIMM(6),AEM(4),ASAM(4),BSAM(4),
     #          AFSAM(6),BFSAM(6),CUTS1(3,4),CUTS2(4,4),CUTS3(4,4),
     #          CUTS4(4,4),CUTS5(4,4)
*
      EXTERNAL D01GCF,REGION,XSC,XSN,XSM43,XSNH19,XSNH24,XSNH49,XSN32,
     #         XSN64,S14AAF,S21BAF,X04CBF,X04ABF,S09ABF
*

\end{verbatim}}\end{quote}

\noindent
This subroutine will decide, according to the input parameters, which
process to consider, e.g. which class and family, and which observable
to compute. For each diagram all the couplings are computed, i.e.
given a current

\begin{equation}
\gamma^{\mu}\,\lb a + b\,\gamma^5\rb ,
\end{equation}

\noindent
the chiral couplings constants

\begin{equation}
\gl = \frac{1}{2}\lb a+b\rb , \qquad \gr = \frac{1}{2}\lb a-b\rb ,
\end{equation}

\noindent
are stored and all the relevant products $\Pi_i\gl^i\Pi_j\gr^j$ are
computed once for all.

All the basic parameters which control the boundaries of
the phase space are initialized in TOW. All the I/O is controlled by this 
subroutine after a call to the subroutine D01GCF which performs the 
integration.
The following functions give the differential cross sections according to
the family to which the process belong. All of them essentially set the
ingredients for the phase space and the requested mappings. Once the vector
{\tt X(NDIM)} is initialized, with $0 \leq x_i \leq 1, i=1,\dots,\,${\tt NDIM},
all independent invariants $u,v,m_-^2,\dots,\ti$ are constructed and from them
all the combinations $x_{ij}, i<j=1,\dots,6$. Next all non-zero helicity
amplitudes are computed and squared. Within these subroutines the operations
among complex numbers are always executed in terms of real and imaginary
parts.

\begin{quote}{\footnotesize \begin{verbatim}

      REAL*8 FUNCTION XSC(NDIM,X)
      IMPLICIT REAL*8 (A-H,O-Z)
      CHARACTER*1,OCOUL,OPEAK,OQCD,OFL
      CHARACTER*10,OTYPE
      DIMENSION X(NDIM)

\end{verbatim}}\end{quote}

\noindent
With {\tt XSC} all the CC3-CC11-CC20 processes are computed. As explained in 
the long write-up this function will allow different choices for the 
observables to be computed, including the total cross section and the moments 
of several distributions. Moreover the variable {\tt OFL}='Y[N]' controls the 
inclusion of the {\it imaginary} part of the $VWW, V=\gamma,Z$ one loop vertex 
in CC20.

\begin{quote}{\footnotesize \begin{verbatim}

      REAL*8 FUNCTION XSNH24(NDIM,X)
      IMPLICIT REAL*8 (A-H,O-Z)
      CHARACTER*1,OQCD
      COMMON/HIGGS/HM
      COMMON/AQCD/OQCD
      DIMENSION X(NDIM)

\end{verbatim}}\end{quote}

\noindent 
With this function the NC25 processes are compute by means of calls to
functions {\tt XSH24} and {\tt XSN}. If {\tt OQCD='Y'} then $\alpha_s(\hm)$ 
is computed through a call to function {\tt RALPHAS} and corrections are 
applied.

\begin{quote}{\footnotesize \begin{verbatim}

      REAL*8 FUNCTION XSNH19(NDIM,X)
      IMPLICIT REAL*8 (A-H,O-Z)
      CHARACTER*1,OQCD
      COMMON/AQCD/OQCD
      DIMENSION X(NDIM)

\end{verbatim}}\end{quote}

\noindent
The same as for function {\tt XSNH24} but for NC21 processes.

\begin{quote}{\footnotesize \begin{verbatim}

      REAL*8 FUNCTION XSNH49(NDIM,X)
      IMPLICIT REAL*8 (A-H,O-Z)
      CHARACTER*1,OQCD
*
      DIMENSION X(NDIM)

\end{verbatim}}\end{quote}

\noindent
The same as for function {\tt XSNH24} but for NC50 processes.

\begin{quote}{\footnotesize \begin{verbatim}

      REAL*8 FUNCTION XSN32(NDIM,X)
      IMPLICIT REAL*8 (A-H,O-Z)
      DIMENSION X(NDIM)

\end{verbatim}}\end{quote}

\noindent
This function will compute all NC32 processes by first calling function {\tt 
XSN}, i.e. the NC24 sub-class (no gluons), and then by calling function {\tt 
XSNG} which will add the remaining $8$ diagrams. This is made possible by the 
fact that there is no interference between NC diagrams and QCD diagrams.
For historical reasons the corresponding Born cross sections ({\tt IOSF}=$0$)
are computed without gluon exchange diagrams for non-leptonic processes.

\begin{quote}{\footnotesize \begin{verbatim}

      REAL*8 FUNCTION XSM43(NDIM,X)
      IMPLICIT REAL*8 (A-H,O-Z)
      CHARACTER*2,OFS
      COMMON/FS/OFS
      DIMENSION X(NDIM)

\end{verbatim}}\end{quote}

\noindent
This function will compute both leptonic and non-leptonic Mix43 processes
by first calling function {\tt XS35}, i.e. the $35$ diagrams without gluons
but with both CC and NC contributions. Next function {\tt XS35G} is called which
includes the QCD diagrams and the interference CC$\otimes$QCD.
For historical reasons the corresponding Born cross sections ({\tt IOSF}=$0$)
are computed without gluon exchange diagrams for non-leptonic processes.

\begin{quote}{\footnotesize \begin{verbatim}

      REAL*8 FUNCTION XSN(NDIM,X)
      IMPLICIT REAL*8 (A-H,O-Z)
      CHARACTER*1,OPEAK
      CHARACTER*10,OTYPE
      COMMON/PS/OPEAK
      COMMON/OCHANNEL/OTYPE
      DIMENSION X(NDIM)

\end{verbatim}}\end{quote}

\noindent
With this function all available NC processes are computed but without
Higgs exchange. Thus it includes the NC19, NC24 and NC48 families.
Only the total cross section is available.

\begin{quote}{\footnotesize \begin{verbatim}

      REAL*8 FUNCTION XSH24(NDIM,X)
      IMPLICIT REAL*8 (A-H,O-Z)
      DIMENSION X(NDIM)

\end{verbatim}}\end{quote}

\noindent
This function will include the Higgs boson exchange diagram to the NC24
processes. It should be stressed that in the limit of massless fermions
there is no interference between NC24 and Higgs so that one can compute
signal and background separately.
If {\tt OQCD}='Y' then $\alpha_s(\wm)$ is evoluted to the $\alpha_s(\hm)$
and the Higgs boson {\it signal} is multiplied by

\begin{eqnarray}
\delta_{_{QCD}} &=& 1 + 5.67\,{{\alpha_s}\over {\pi}} + 
42.74\,\lb {{\alpha_s}\over {\pi}}\rb^2,  \nll
\alpha_s &=& \alpha_s(\hm).
\end{eqnarray}

\begin{quote}{\footnotesize \begin{verbatim}

      REAL*8 FUNCTION XSH19(NDIM,X)
      IMPLICIT REAL*8 (A-H,O-Z)
      DIMENSION X(NDIM)

\end{verbatim}}\end{quote}

\noindent
With this function the Higgs boson exchange is added to the NC19 family
thus obtaining NC21. The two additional diagrams again do not interfere
with NC19 in the massless limit.

\begin{quote}{\footnotesize \begin{verbatim}

      REAL*8 FUNCTION XSNH49(NDIM,X)
      IMPLICIT REAL*8 (A-H,O-Z)
      CHARACTER*1,OQCD
*
      COMMON/QCD/ALS
      COMMON/HIGGS/HM
      COMMON/AQCD/OQCD
*
      DIMENSION X(NDIM)

\end{verbatim}}\end{quote}

\noindent
With this function the Higgs boson exchange is added to the NC48 family
thus obtaining NC50. The two additional diagrams again do not interfere
with NC48 in the massless limit.

\begin{quote}{\footnotesize \begin{verbatim}

      REAL*8 FUNCTION XSNG(NDIM,X)
      IMPLICIT REAL*8 (A-H,O-Z)
      CHARACTER*10,OTYPE
      COMMON/OCHANNEL/OTYPE
      DIMENSION X(NDIM)

\end{verbatim}}\end{quote}

\noindent
The gluon exchange diagrams for non-leptonic NC processes are included.
As explained in the previous sections we use a {\it fixed} $\alpha_s$,
i.e. $\alpha_s(\wm)$.

\begin{quote}{\footnotesize \begin{verbatim}

      REAL*8 FUNCTION XS35(NDIM,X)
      IMPLICIT REAL*8 (A-H,O-Z)
      CHARACTER*1,OCOUL,OPEAK
      CHARACTER*2,OFS
      CHARACTER*10,OTYPE
      COMMON/FS/OFS
      COMMON/PS/OPEAK
      COMMON/COUL/OCOUL
      COMMON/OCHANNEL/OTYPE
      DIMENSION X(NDIM)

\end{verbatim}}\end{quote}

\noindent
This function computes the diagrams for CC$\otimes$NC leptonic processes
and for the non-leptonic ones where gluon exchange will be included
separately.

\begin{quote}{\footnotesize \begin{verbatim}

      REAL*8 FUNCTION XS35G(NDIM,X)
      IMPLICIT REAL*8 (A-H,O-Z)
      CHARACTER*1,OCOUL,OPEAK
      CHARACTER*2,OFS
      CHARACTER*10,OTYPE
      COMMON/FS/OFS
      COMMON/PS/OPEAK
      COMMON/COUL/OCOUL
      COMMON/OCHANNEL/OTYPE
      DIMENSION X(NDIM)

\end{verbatim}}\end{quote}
 
\noindent
Whenever CC$\otimes$NC non-leptonic processes are to be computed this
function will return QCD$\otimes$(QCD+CC).

\begin{quote}{\footnotesize \begin{verbatim}

      REAL*8 FUNCTION XSN64(NDIM,X)
      IMPLICIT REAL*8 (A-H,O-Z)
      CHARACTER*1,OPEAK
      CHARACTER*2,OFS
      COMMON/FS/OFS
      COMMON/PS/OPEAK
      DIMENSION X(NDIM)
\end{verbatim}}\end{quote}

\noindent
Finally this function allows to compute processes with identical particles in 
the final state, with the exclusion of $e^{\pm}$ and of $\nu_e$. For the
non-leptonic case QCD is included and the proper color factor relative to
each diagram is automatically computed.

\begin{quote}{\footnotesize \begin{verbatim}

      SUBROUTINE REGION(N,X,J,A,B)
      IMPLICIT REAL*8(A-H,O-Z)
      DIMENSION X(N)

\end{verbatim}}\end{quote}

\noindent
This subroutine is requested by D01GCF in order to set the limits of 
integrations. In WTO its content is trivial since all variables are in the
range $[0,1]$.

\begin{quote}{\footnotesize \begin{verbatim}

      SUBROUTINE CORRQCD(SCAL,ALS,BQM2,CQM2,RBM2,RCM2)
      IMPLICIT REAL*8(A-H,O-Z)
      COMMON/BPAR/WM,ZM,ZG,GF,EM,PI,CFCT,FCNT,GE,TM,BQM,CQM,ALWI
      DATA Z3/1.20206D0/,NF/5/

\end{verbatim}}\end{quote}

\noindent
Computes the QCD quark running masses. Given {\tt ALS}, the value of 
$\alpha_s(\wm)$, and the pole $b,c$-quark masses this routine evaluates 
their running at a scale $\mu =\,${\tt SCAL}.

\begin{quote}{\footnotesize \begin{verbatim}

      REAL*8 FUNCTION RALPHAS(RS0,RS,ALS,NF)
      IMPLICIT REAL*8(A-H,O-Z)
      COMMON/BPAR/WM,ZM,ZG,GF,EM,PI,CFCT,FCNT,GE,TM,BQM,CQM,ALWI

\end{verbatim}}\end{quote}

\noindent
Given {\tt ALS}, the value of $\alpha_s$ at a scale {\tt RS0}, this routine 
returns $\alpha_s$ at the scale $\mu =\,${\tt RS} with {\tt NF} active flavors.

\begin{quote}{\footnotesize \begin{verbatim}

      REAL*8 FUNCTION QCDLAM(NF,ALS,RS,X1,X2,XACC)
      IMPLICIT REAL*8(A-H,O-Z)
      PARAMETER (JMAX=50,NOUT=21)

      REAL*8 FUNCTION QCDSCALE(NF,ALS,RS,X)
      IMPLICIT REAL*8(A-H,O-Z)

\end{verbatim}}\end{quote}

\noindent
Compute $\Lambda^{n_f}$ in the $\msb$-scheme from $\alpha_s$ fixed
at the scale {\tt RS0}.

\begin{quote}{\footnotesize \begin{verbatim}

      SUBROUTINE HADR5(E,DER,EDER)
******************************************************************
*                                                                *
*      SUBROUTINE FOR THE EVALUATION OF THE LIGHT HADRON         *
*           CONTRIBUTIONS TO DELTA_R                             *
*                                                                *
*    F. JEGERLEHNER, PAUL SCHERRER INSTITUTE, CH-5232 VILLIGEN   *
*                                                                *
*    REFERENCE: F. JEGERLEHNER, Z. PHYS. C32 (1986) 195          *
*               H. BURKHARDT ET AL., Z. PHYS. C42 (1989) 497     *
*               S. EIDELMAN, F. JEGERLEHNER, Z. PHYS. C (1995)   *
*                                                                *
******************************************************************
*       VERSION: 24/02/1995
*
*  THIS PROGRAM DOES NOT YET KNOW HOW TO COMPUTE DELTA R AND DELTA G FOR
*  ENERGIES IN THE RANGES  |E|>1TEV AND 2M_PI < E < 40(13) GEV !!!!!!!!!
*
      IMPLICIT REAL*8(A-H,O-Z)

\end{verbatim}}\end{quote}

\begin{quote}{\footnotesize \begin{verbatim}

      SUBROUTINE PSELF(P2X,PGGF)
      IMPLICIT REAL*8(A-H,O-Z)
*
      COMMON/PARAM/EPS,QDELTA
      COMMON/FMASS/EM,RMM,TM,TQM,UQM,DQM,CQM,SQM,BQM
      COMMON/BPAR/WM,ZM,ZG,GF,PI,CFCT,FCNT,GE,ALPHAI,ALWI

\end{verbatim}}\end{quote}

\noindent 
{\tt PSELF} computes the {\it perturbative} contribution to the running of
$\alpha(s)$.

\begin{quote}{\footnotesize \begin{verbatim}

      SUBROUTINE RBFF0(RM12,RM22,B0,B1,B21)
      IMPLICIT REAL*8(A-H,O-Z)
*
      COMMON/PARAM/EPS,QDELTA
      COMMON/BPAR/WM,ZM,ZG,GF,PI,CFCT,FCNT,GE,ALPHAI,ALWI
*
      DIMENSION ARG(2),CLN(2),FR(3)

      SUBROUTINE RBFF(P2,RM12,RM22,RB0,RB1,RB21)
      IMPLICIT REAL*8(A-H,O-Z)
*
      COMMON/PARAM/EPS,QDELTA
      COMMON/BPAR/WM,ZM,ZG,GF,PI,CFCT,FCNT,GE,ALPHAI,ALWI
*
      DIMENSION CMP2(2)
      DIMENSION CLNMP2(2)
      DIMENSION GFPR(3),GFMR(3)

      SUBROUTINE ROOTS(P2,RM12,RM22,RPR,RPI,RMR,RMI,OMRPR,OMRMR)
      IMPLICIT REAL*8(A-H,O-Z)
*
      COMMON/PARAM/EPS,QDELTA
      COMMON/BPAR/WM,ZM,ZG,GF,PI,CFCT,FCNT,GE,ALPHAI,ALWI

      SUBROUTINE CQLNX(ARG,RES)
      IMPLICIT REAL*8(A-H,O-Z)
*
      COMMON/BPAR/WM,ZM,ZG,GF,PI,CFCT,FCNT,GE,ALPHAI,ALWI
*
      DIMENSION ARG(2),AARG(2),RES(2)

      SUBROUTINE CQLNOMX(ARG,OMARG,RES)
      IMPLICIT REAL*8(A-H,O-Z)
*
      DIMENSION ARG(2),OMARG(2),RES(2),ARES(2),CT(10),SN(10)

      SUBROUTINE RCG(N,ZR,ZI,OMZR,GFR)
      IMPLICIT REAL*8(A-H,O-Z)
*
      COMMON/PARAM/EPS,QDELTA
      COMMON/BPAR/WM,ZM,ZG,GF,PI,CFCT,FCNT,GE,ALPHAI,ALWI
*
      DIMENSION GFR(N)
      DIMENSION A(4),RA(16),Z(2),OMZ(2),OZ(2),CLNOMZ(2),CLNOZ(2)
      DIMENSION CA(2,10),AUX(2),ZP(2),CT(16),SN(16)

\end{verbatim}}\end{quote}

\noindent
These routines are required by {\tt PSELF} and compute the relevant 
{\it perturbative} contributions to the photon two point function.

\section{Test run output}

\subsection{Sample 1}

\begin{quote}{\footnotesize \begin{verbatim}

This run is with: 

NPTS         =  8
NRAND        =  6

1/alpha_QED(s) = 0.12807E+03 Coupling constants are (alpha-G_F) : 

gv_e    =      -0.1409725455532E-01     -0.1409725455532E-01
ga_e    =      -0.1857939837526E+00     -0.1857939837526E+00
g_wf    =       0.2304098503535E+00      0.2304098503535E+00
g_zww   =      -0.5714792058132E+00     -0.5714792058132E+00

G-L factor  =     0.104114673716395E+01

E_cm (GeV) =          0.19000E+03
beta       =          0.11453E+00 sin^2     =          0.23103E+00
M_W  (GeV) =          0.80230E+02 M_Z (GeV) =          0.91189E+02
G_W  (GeV) =          0.20337E+01 G_Z (GeV) =          0.24974E+01

alpha RS
O(alpha^2) beta SF 
CS not included
FSR not included
NQCD not included

Process is e+ e- -> mu-     bnu_mu     u    bd   
with the following cuts 

 E(GeV), SA(deg) cuts

              mu-     bnu_mu          u         bd
 E_th      1.0000     0.0000     3.0000     3.0000
C_max     10.0000     0.0000     0.0000     0.0000
C_min    170.0000   180.0000   180.0000   180.0000

 lower IM cuts (GeV)

              mu-    bnu_mu       u      bd
   mu-               0.0000  0.0000  0.0000
bnu_mu                       0.0000  0.0000
     u                               5.0000
    bd

 upper IM cuts (GeV)

              mu-     bnu_mu          u         bd
   mu-              190.0000   190.0000   190.0000
bnu_mu                         190.0000   190.0000
     u                                    190.0000
    bd

 min FS angle(deg) cuts

              mu-     bnu_mu          u         bd
   mu-              180.0000   180.0000   180.0000
bnu_mu                         180.0000   180.0000
     u                                    180.0000
    bd

 max FS angle(deg) cuts

              mu-     bnu_mu          u         bd
   mu-                0.0000     5.0000     5.0000
bnu_mu                           0.0000     0.0000
     u                                      0.0000

cc11-diagrams : charges    -1.0000    0.0000    0.6667    0.3333

Cross-Section On exit IFAIL = 0

CPU time   4 h  15 min   5 sec

sec per call = 0.249E-02

sigma = 0.5919248E+00 +- 0.4040678E-04

Rel. error of      0.007 %

\end{verbatim}}\end{quote}

\subsection{Sample 2}

\begin{quote}{\footnotesize \begin{verbatim}

This run is with: 

NPTS         = 10
NRAND        =  5

1/alpha_QED(s) = 0.12807E+03 Coupling constants are (alpha-G_F) : 

gv_e    =      -0.1409725455532E-01     -0.1409725455532E-01
ga_e    =      -0.1857939837526E+00     -0.1857939837526E+00
g_wf    =       0.2304098503535E+00      0.2304098503535E+00
g_zww   =      -0.5714792058132E+00     -0.5714792058132E+00

G-L factor  =     0.104114673716395E+01

E_cm (GeV) =          0.19000E+03
beta       =          0.11453E+00 sin^2     =          0.23103E+00
M_W  (GeV) =          0.80230E+02 M_Z (GeV) =          0.91189E+02
G_W  (GeV) =          0.20337E+01 G_Z (GeV) =          0.24974E+01

alpha RS
No QED Radiation 

Process is e+ e- -> d       bd      e-      e+     
with the following cuts 

 E(GeV), SA(deg) cuts
                d         bd         e-         e+
 E_th      3.0000     3.0000     1.0000     1.0000
C_max      0.0000     0.0000    10.0000    10.0000
C_min    180.0000   180.0000   170.0000   170.0000

 lower IM cuts (GeV)
             d         bd         e-         e+
 d                 5.0000     0.0000     0.0000
bd                            0.0000     0.0000
e-                                       0.0000
e+

 upper IM cuts (GeV)
             d         bd         e-         e+
 d               190.0000   190.0000   190.0000
bd                          190.0000   190.0000
e-                                     190.0000
e+

 min FS angle(deg) cuts
             d         bd         e-         e+
 d               180.0000   180.0000   180.0000
bd                          180.0000   180.0000
e-                                     180.0000
e+

 max FS angle(deg) cuts
             d         bd         e-         e+
 d                 0.0000     5.0000     5.0000
bd                            5.0000     5.0000
e-                                       5.0000
e+

nc48-diagrams : charges    -0.3333   -1.0000
                isospin    -0.5000   -0.5000

Cross-Section On exit IFAIL = 0

CPU time   9 h   9 min   3 sec

sec per call = 0.111E-02

sigma = 0.4304695E-01 +- 0.5671371E-04

Rel. error of      0.132 %

\end{verbatim}}\end{quote}

\subsection{Sample 3}

\begin{quote}{\footnotesize \begin{verbatim}

This run is with: 

NPTS         =  9
NRAND        =  6

1/alpha_QED(s) = 0.12807E+03 Coupling constants are (alpha-G_F) : 

gv_e    =      -0.1409725455532E-01     -0.1409725455532E-01
ga_e    =      -0.1857939837526E+00     -0.1857939837526E+00
g_wf    =       0.2304098503535E+00      0.2304098503535E+00
g_zww   =      -0.5714792058132E+00     -0.5714792058132E+00

G-L factor  =     0.104114673716395E+01

E_cm (GeV) =          0.19000E+03
beta       =          0.11453E+00 sin^2     =          0.23103E+00
M_W  (GeV) =          0.80230E+02 M_Z (GeV) =          0.91189E+02
G_W  (GeV) =          0.20337E+01 G_Z (GeV) =          0.24974E+01

alpha RS
O(alpha^2) beta SF 
CS not included

Process is e+ e- -> mu-     bnu_mu  nu_mu   mu+    
with the following cuts 

 E(GeV), SA(deg) cuts
              mu-     bnu_mu      nu_mu        mu+
 E_th      1.0000     0.0000     0.0000     1.0000
C_max     10.0000     0.0000     0.0000    10.0000
C_min    170.0000   180.0000   180.0000   170.0000

 lower IM cuts (GeV)
            mu-  bnu_mu   nu_mu     mu+
   mu-           0.0000  0.0000  0.0000
bnu_mu                   0.0000  0.0000
 nu_mu                           0.0000
   mu+

 upper IM cuts (GeV)
               mu-     bnu_mu      nu_mu        mu+
   mu-               190.0000   190.0000   190.0000
bnu_mu                          190.0000   190.0000
 nu_mu                                     190.0000
   mu+

 min FS angle(deg) cuts
               mu-     bnu_mu      nu_mu        mu+
   mu-               180.0000   180.0000   180.0000
bnu_mu                          180.0000   180.0000
 nu_mu                                     180.0000
   mu+

 max FS angle(deg) cuts
               mu-     bnu_mu      nu_mu        mu+
   mu-                 0.0000     0.0000     5.0000
bnu_mu                            0.0000     0.0000
 nu_mu                                       0.0000
   mu+

Cross-Section On exit IFAIL = 0

CPU time   7 h  59 min   2 sec

sec per call = 0.139E-02

sigma = 0.2046758E+00 +- 0.1704597E-03

Rel. error of      0.083 %

\end{verbatim}}\end{quote}

\section*{Acknowledgements}
%\addcontentsline{toc}{section}{Acknowledgements}

The WTO program is a result of a long term project which started with 
an intense and fruitful collaboration between Torino and Pavia. Here
I acknowledge many important discussions with Oreste~Nicrosini, 
Guido~Montagna and expecially with Fulvio~Piccinini. In the development
of the program I had the unique opportunity of a day by day collaboration
with Dima~Bardin. It is also because of the continuous interaction
between the two of us that WTO can now make its appearance. In particular
I sincerely acknowledge the active contribute of Dima~Bardin in the
comparison phase with GENTLE/4-fan where, expecially for CC11, we have been 
able to reach agreement with very high precision results. I also 
acknowledge many discussions with Roberto~Pittau and comparisons with
EXCALIBUR. {\it Lively} discussions with Ronald~Kleiss and more generally with
all the members of the WW/eg working group for LEP~2 physics are acknowledged.
This list would not be complete without acknowledging many friends, both 
theorists and experimentalists, who have been so generous to give me full 
support with their computing resources.
Finally I must acknowledge that the constant collaboration
of Alessandro~Ballestrero was an essential ingredient towards the completion 
of WTO. The continuous exchange of ideas, informations, results between
WTO and its companion WPHACT has been the main guidance through several 
months of work.

\twocolumn
\section{Appendix}
\vskip 25pt

\begin{enumerate}

\begin{itemize}
\item CC11
\end{itemize}

\item $      \mu^-, \barnu _{\mu}, \nu_{\tau}, \tau^+$
\item $      \mu^-, \barnu _{\mu}, u, {\bar d}$
\item $      d, {\bar u}, c, {\bar s}$

\begin{itemize}
\item CC20
\end{itemize}

\item $      e^-, \barnu _e, \nu_{\mu}, \mu^+$
\item $      e^-, \barnu _e, u, {\bar d}$

\begin{itemize}
\item NC24
\end{itemize}

\item $      \mu^-, \mu^+, \nu_{\tau}, \barnu _{\tau}$
\item $      d, {\bar d}, \nu_{\mu}, \barnu _{\mu}$
\item $      u, {\bar u}, \nu_{\mu}, \barnu _{\mu}$
\item $      \mu^-, \mu^+, \tau^-, \tau^+$
\item $      \mu^-, \mu^+, d, {\bar d}$
\item $      \mu^-, \mu^+, u, {\bar u}$
\item $      \nu_{\mu}, \barnu _{\mu}, \nu_{\tau}, \barnu _{\tau}$

\begin{itemize}
\item NC32
\end{itemize}

\item $      s, {\bar s}, u, {\bar u}$
\item $      d, {\bar d}, s, {\bar s}$
\item $      u, {\bar u}, c, {\bar c}$

\begin{itemize}
\item NC19
\end{itemize}

\item $      \nu_{\mu}, \barnu _{\mu}, \nu_e, \barnu _e$
\item $      \mu^-, \mu^+, \nu_e, \barnu _e$
\item $      u, {\bar u}, \nu_e, \barnu _e$
\item $      d, {\bar d}, \nu_e, \barnu _e$

\begin{itemize}
\item Mix43
\end{itemize}

\item $      \mu^-, \barnu _{\mu}, \nu_{\mu}, \mu^+$
\item $      d, {\bar u}, u, {\bar d}$

\begin{itemize}
\item NC48
\end{itemize}

\item $      \mu^-, \mu^+, e^-, e^+$
\item $      \nu_{\mu}, \barnu _{\mu}, e^-, e^+$
\item $      u, {\bar u}, e^-, e^+$
\item $      d, {\bar d}, e^-, e^+$

\begin{itemize}
\item NC64
\end{itemize}

\item $      \mu^-, \mu^+, \mu^-, \mu^+$
\item $      \nu_{\mu}, \barnu _{\mu}, \nu_{\mu}, \barnu _{\mu}$
\item $      u, {\bar u}, u, {\bar u}$
\item $      d, {\bar d}, d, {\bar d}$

\begin{itemize}
\item NC25(NC24+1)
\end{itemize}

\item $      b, \barb , \nu_{\mu}, \barnu _{\mu}$
\item $      \mu^-, \mu^+, b, \barb $

\begin{itemize}
\item NC21(NC19+1)
\end{itemize}

\item $      b, \barb , \nu_e, \barnu_e$

\begin{itemize}
\item NC50(NC48+2)
\end{itemize}

\item $      b, \barb, e^-, e^+$
\end{enumerate}

\onecolumn

\newpage

%===============================================================================


\begin{thebibliography}{99}
%===============================================================================

\addcontentsline{toc}{section}{References}

\bibitem{nag}  NAG Fortran Library Manual Mark 15 (Numerical Algorithms Group),
Oxford, 1991.

\bibitem{mhf} G.~Passarino, Nucl. Phys. B237(1984)249.

\bibitem{wweg} Report on {\it Event Generators for WW Physics}
in {\it Proceedings of the CERN Workshop on Physics at LEP2}, CERN
Yellow Report, to appear in 1996.

\bibitem{lep2}
D.~Bardin and T.~Riemann, preprint DESY~95--167 (1995) [hep-ph/9509341];\\
E. Boos et al., Moscow State Univ. preprint MGU-89-63/140 (1989);\\
preprint KEK 92-47 (1992);\\
D.~Bardin, M.~Bi\-len\-ky, D.~Leh\-ner, A.~Ol\-chev\-ski and T.~Riemann,
in: T.~Riemann and J.~Bl\"umlein (eds.),
Proc. of the Zeuthen Workshop on Elementary Particle Theory --
Physics at LEP200 and Beyond, Teupitz, Germany,
April 10--15, 1994, Nucl. Phys. (Proc. Suppl.) 37B(1994) p. 148;\\
D. Bardin, A. Leike and T. Riemann, Phys. Lett. B353(1995)513;\\
F.A.~Berends, R.~Kleiss and R.~Pittau,
Comp.Phys.Comm. 85(1995)437;\\
Nucl.Phys. B424(1994)308;\\
Nucl.Phys. B426(1994)344;\\
Nucl.Phys. B, Proc. Suppl. 37B(1994)163;\\
R.~Pittau, Phys. Lett. B335(1994)490;\\
F. Caravaglios and M. Moretti, Phys. Lett. B358(1995)332;\\
C.G.~Papadopoulos, Phys.~Lett. B352(1995)144;\\
D. Bardin, M. Bilenky, A.~Olchevski and T.~Riemann,
Phys. Lett. B308(1993)403; E:[hep-ph/9507277];\\
Minami-Tateya collaboration, KEK Report 92-19,1993; \\
S. Kawabata, Comp. Phys. Commun.  88(1995)309;\\
M. Skrzypek, S. Jadach, W. P\l{}aczek, and Z. W\c{a}s, 
CERN preprint CERN-TH/95-205;\\
T.~Sj\"ostrand, Comp. Phys. Commun. 82(1994)74;\\
H.~Anlauf et al., Comp.~Phys.~Comm. 79(1994)487 ;\\
E.~Accomando and A.~Ballestrero, Fortran program WPHACT, in preparation;\\
G.~Montagna, O.~Nicrosini, G.~Passarino and F.~Piccinini,
Phys. Lett. B348(1995)178;\\
G. J. van Oldenborgh, P. J. Franzini, and A. Borrelli,  Comp. Phys.
Comm.  83(1994)14;\\
G.~Montagna, O.~Nicrosini and F.~Piccinini, Comp. Phys. Commun.
90(1995)141;\\

\bibitem{yr} G.~Altarelli et al. eds. Proceedings of the Workshop on {\it
Physics at LEP 2} CERN Yellow Report in preparation.

\bibitem{vwg} Report on {\it Standard Model Processes}
in {\it Proceedings of the CERN Workshop on Physics at LEP2}, CERN
Yellow Report, to appear in 1996; \\
Report on {\it Event Generators for Discovery Physics}, ibid.

\bibitem{jeg} F.~Jegerlehner, private communication.

\bibitem{sf1} 
G.~Montagna, O.~Nicrosini and F.~Piccinini, Comp. Phys. Commun.
90(1995)141;\\

\bibitem{sf2}
G.~Montagna, O.~Nicrosini, G.~Passarino and F.~Piccinini,
Phys. Lett. B348(1995)178.

\bibitem{sf3}
F.~A.~Berends, R.~Kleiss and R.~Pittau, Nucl. Phys. 
B426(1994)344.

\bibitem{amp} P.~DeCausmaeker, R.~Gastmans, W.~Troost and T.~T.~Wu,
Nucl. Phus. B206(1982)53; \\
M.~Caffo and E.~Remiddi, Helv. Phys. Acta 55(1982)339;  \\
G.~Passarino, Phys. Rv. D28(1983)2867.

\bibitem{gi}
E.~N.~Argyres et al., Phys Lett. {\bf B 358}(1995)339.

\bibitem{gen} D.~Bardin, M.~Bi\-len\-ky, D.~Leh\-ner, A.~Ol\-chev\-ski and T.~Riemann,
in: T.~Riemann and J.~Bl\"umlein (eds.),
Proc. of the Zeuthen Workshop on Elementary Particle Theory --
Physics at LEP200 and Beyond, Teupitz, Germany,
April 10--15, 1994, Nucl. Phys. (Proc. Suppl.) 37B(1994) p. 148.

\bibitem{cgr} M.~Caffo, R.~Gatto and E.~Remiddi, Nucl. Phys. B252(1985)378.

\bibitem{cmn} M.~Cacciari, G.~Montagna and O.~Nicrosini, Phys. Lett. 
B274(1992)473.

%-------------------------------------------------------------------

\end{thebibliography}
\end{document}